\documentclass[12pt, hyper]{JHEP3}  
\usepackage{amsmath,amssymb,amsfonts,amsxtra, mathrsfs, makeidx,graphics,graphicx,amsthm,epsfig}
\pdfoutput=1

\newcommand{\be}{\begin{equation}}
\newcommand{\ee}{\end{equation}}
\newcommand{\beq}{\begin{equation}}
\newcommand{\eeq}{\end{equation}}
\newcommand{\ba}{\begin{array}}
\newcommand{\ea}{\end{array}}
\newcommand{\bi}{\begin{itemize}}
\newcommand{\ei}{\end{itemize}}
\newcommand{\ben}{\begin{enumerate}}
\newcommand{\een}{\end{enumerate}}
\newcommand{\bea}{\begin{eqnarray}}
\newcommand{\eea}{\end{eqnarray}}
\newcommand{\bean}{\begin{eqnarray*}}
\newcommand{\eean}{\end{eqnarray*}}
\newcommand{\eref}[1]{(\ref{#1})}
\newcommand{\sref}[1]{\S\ref{#1}}
\newcommand{\tref}[1]{Table~\ref{#1}}
\newcommand{\fref}[1]{Figure~\ref{#1}}
\newcommand{\nn}{\nonumber}

\newcommand{\tr}{\mathop{\rm Tr}}

\newcommand{\BC}{\mathbb{C}}
\newcommand{\BR}{\mathbb{R}}

\newcommand{\BZ}{\mathbb{Z}}

\newcommand{\cN}{{\cal N}}

\def\tQ{\widetilde{Q}}
\def\tB{\widetilde{B}}

\newcommand{\PE}{\mathrm{PE}}
\newcommand{\PL}{\mathrm{PL}}

\newcommand{\ud}{\mathrm{d}}

\newcommand{\comment}[1]{}

\newcommand{\f}{{\cal F}^{\flat}}
\newcommand{\gf}{g^{\f}}

\newcommand{\MH}{{\cal M}^{\mathrm{Higgs}}}
\newcommand{\gH}{{g}^{\mathrm{Higgs}}}

\newcommand{\mN}{\mathcal{N}}
\newcommand{\HS}{g^\mathrm{Irr}}
\newcommand{\qt}{\tilde{q}}

\newcommand{\eg}{\emph{e.g.}}

 \title{The Hilbert Series of the One Instanton Moduli Space}
\author{Sergio Benvenuti, Amihay Hanany and Noppadol Mekareeya\\
Theoretical Physics Group, The Blackett Laboratory \\
Imperial College London, Prince Consort Road\\ 
London,  SW7 2AZ,  UK \\
Email: {\tt s.benvenuti, a.hanany, n.mekareeya07@imperial.ac.uk}}
\abstract{The moduli space of $k$ $G$-instantons on $\BR^4$ for a classical gauge group $G$ is known to be given by  the Higgs branch of a supersymmetric gauge theory that lives on D$p$ branes probing D$(p + 4)$ branes in Type II theories. For $p$ = 3, these (3 + 1) dimensional gauge theories have $\mN = 2$ supersymmetry and can be represented by quiver diagrams. The F and D term equations coincide with the ADHM construction. The Hilbert series of the moduli spaces of one instanton for classical gauge groups is easy to compute and turns out to take a particularly simple form which is previously unknown. This allows for a $G$ invariant character expansion and hence easily generalisable for exceptional gauge groups, where an ADHM construction is not known. The conjectures for exceptional groups are further checked using some new techniques like sewing relations in Hilbert Series. This is applied to Argyres-Seiberg dualities.
}

\begin{document}

\section{Introduction}
Yang-Mills Instantons \cite{Belavin:1975fg} have attracted great interest from both physicists and mathematicians since their discovery in 1975.  They have served as a powerful tool in studying a number of physical and mathematical problems, ranging from the Yang-Mills vacuum structure (\eg, \cite{thooft, jacreb, cdg}) to the classification of four-manifolds \cite{don}.  

A method for constructing a self-dual Yang-Mills instanton solution on $\BR^4$ is due to Atiyah, Drinfeld, Hitchin and Manin (ADHM) \cite{Atiyah:1978ri} in 1978.  The ADHM construction is known for the classical gauge groups, $SU(N)$, $SO(N)$ and $Sp(N)$ (see, \eg, \cite{christ, Corrigan:1983sv, dkm, Nekrasov:2004vw, Marino:2004cn} for explicit constructions); there is no known such construction, however, for the exceptional groups.  The space of all solutions to the self-dual Yang-Mills equation modulo gauge transformations, in a given winding sector $k$ and gauge group $G$ is said to be the {\bf moduli space of $k$ $G$-instantons} on $\BR^4$.  In 1994-1996, Douglas and Witten \cite{Witten:1994tz, Douglas:1995bn, Douglas:1996uz, Witten:1995gx} discovered that the ADHM construction can be realised in string theory.  In particular, the moduli space of instantons on $\BR^4$ is identical to the Higgs branch of supersymmetric gauge theories on a system of D$p$-D$(p+4)$ branes (see, \eg, \cite{Tong:2005un} for a review).\footnote{The Higgs branch of D3 branes near $E_n$ type 7 branes is the moduli space of $E_n$ instantons. Since there is no known Lagrangian for this class of theories, it is not clear how to compute the ADHM analog.}   These theories are quiver gauge theories with 8 supercharges ($\cN=2$ supersymmetry in $(3+1)$ dimensions for $p=3$).  In \sref{classical} of this paper, we present the $\cN=2$ quiver diagram of each theory as well as provide a prescription for writing down the corresponding $\cN=1$ quiver diagram and the superpotential.  The Hilbert series of the one instanton moduli space is easily computed using the ADHM construction for classical gauge groups and is written in a form that provides a natural conjectured generalization for exceptional gauge groups (even though the ADHM construction does not exist for the latter).

In addition to the ADHM construction, there exists an alternative description of the moduli space of instantons for simply laced ($A$, $D$ and $E$) groups via three dimensional mirror symmetry \cite{Intriligator:1996ex}.  This symmetry exchanges the Coulomb branch and the Higgs branch, and therefore maps the Coulomb branch of the $ADE$ quiver gauge theories to moduli spaces of instantons.  On the contrary to Higgs branch, one expects the Coulomb branch to receive many non-perturbative quantum corrections.  As argued in \cite{Intriligator:1996ex}, quantum effects correct the Coulomb branch to be the moduli space of one $ADE$-instanton, with the point at the origin corresponding to an instanton of zero size.\footnote{The Coulomb branch of the gauge theory with quiver diagram $G$ (where $G$ is $A$, $D$ or $E$) and all ranks multiplied by $k$ is $kh_G-1$ quaternionic dimensional \cite{Intriligator:1996ex}, where $h_G$ is the dual coxeter number of $G$.  This precisely agrees with the fact that the coherent component (eliminating the translation on $\BR^4$) of the one $G$-instanton moduli space is $h_G-1$ quaternionic dimensional.}  Nevertheless, due to such quantum corrections, this description of the instanton moduli space is not useful for exact computations using Hilbert series.

In the last section of this paper, exceptional groups are considered, and checks that the Hilbert Series above predicts the correct dimension of the moduli space. In the case of $E_n$ it is known \cite{Minahan:1996fg, Minahan:1996cj} that $\mN =2$ CFTs realise the moduli space of one $E_n$ instanton. We use Argyres-Seiberg S-dualities in $\cN=2$ supersymmetric gauge theories  \cite{Argyres:2007cn, Argyres:2007tq, Gaiotto:2008nz, Gaiotto:2009we, Gaiotto:2009gz, Tachikawa:2009rb,  Benini:2009gi} to match the Hilbert series of the theories on both sides of the duality, providing a consistency check.


\section{Hilbert series for one-instanton moduli spaces on $\BC^2$}
We are interested in computing the partition function that counts holomorphic functions (Hilbert series) on the moduli space of $k$ $G$-instantons on $\BC^2$, were $G$ is a gauge group of finite rank $r$. It is well known that this moduli space has quaternionic dimension $kh_G$ where $h_G$ is the dual Coxeter number of the gauge group $G$. The present paper will focus on the case of a single instanton moduli space. The moduli space is reducible into a trivial $\BC^2$ component, physically corresponding to the position of the instanton in $\BC^2$, and the remaining irreducible component of quaternionic dimension $h_G-1$. Henceforth, we shall call this component the {\it coherent component} or the {\it irreducible component}.  The Hilbert series for the coherent component takes the form
\bea 
\HS_G (t; x_1,\ldots,x_r) = \sum_{k=0}^\infty \chi[ R_G(k) ] t^{2k}~, \label{HS}
\eea
where $R_G (k)$ is a series of representations of $G$ and $\chi[R]$ is the character of the representation $R$.\footnote{In this paper, we represent an irreducible representation of a group G by its Dynkin labels (which is also the highest weight of such a representation) $[a_1,...,a_r]$, where $r=\mathrm{rank}~G$.  Since a representation is determined by its character, we slightly abuse terminology by referring to a character by the corresponding representation.} The fugacities $x_i$  (with $i=1,\ldots, r$) are conjugate to the charges of each holomorphic function under the Cartan subalgebra of $G$.  The moduli space of instantons is a non-compact hyperK\"ahler space, and so there are infinitely many holomorphic functions which are graded by degrees $d$. 
Setting $x_1=\ldots = x_r =1$, we obtain the (finite) number of holomorphic functions of degree $d$.


The main result of this paper is the following:
\begin{quote}
\emph{The representation $R_G(k)$ is the irreducible representation $Adj^k$~,}
\end{quote}
where $Adj^k$ denotes the irreducible 
representation whose Dynkin labels are $\theta_k = k \theta$, with $\theta$ the highest root  of $G$.\footnote{For the $A_n$ series $\theta_k = [k,0,\ldots,0,k]$, for the $B_n$ and $D_n$ series $\theta_k=[0,k,0,\ldots,0]$, for the $C_n$ series $\theta_k=[2k,0,\ldots,0]$, for $E_6$ $\theta_k = [0,k,0,0,0,0]$, for $G_2$ $\theta_k = [0,k]$, for all other exceptional groups $\theta_k = [k,0,\ldots,0]$.} By convention $R_G (0)$ is the trivial, one-dimensional, representation (this corresponds to the space being connected), and $R_G(1)$ is the adjoint representation.

In the case of classical gauge groups $A_n, B_n, C_n, D_n$ it is possible to directly verify the above statement by explicit counting of the chiral operators on the Higgs branch of a certain $\mN=2$ supersymmetric gauge theory with a one dimensional Coulomb branch and a $A_n, \ldots, D_n$ global symmetry. The specific gauge theory can be derived in string theory by a simple system of D$p$ branes which probe a background of D$(p+4)$ branes in Type II theories. The moduli space of $k$ $G$-instantons on $\BC^2$ is identified with the Higgs branch of the gauge theory living on the $k$ D$p$ branes. The gauge group $G$, which is interpreted as a global symmetry on the world volume of the D$p$ branes, lives on the D$(p+4)$ branes and can be chosen to be any of the classical gauge groups by an appropriate choice of a background with or without an orientifold plane. The gauge theory living on the D$p$ branes is a simple quiver gauge theory and is discussed in detail in \sref{classical}. 
The F and D term equations for the Higgs branch of these theories coincides with the ADHM construction of the moduli space of instantons for classical gauge groups. Unfortunately, such a simple construction is not available for exceptional groups and other methods need to be applied.
It is therefore not possible to explicitly compute the Hilbert series for exceptional groups and the main statement of this paper is a conjecture for these cases. This conjecture is subject to a collection of tests which are presented in \sref{EN}.

\paragraph{An example of $D_4$.}
An explicit counting of chiral operators in the well known $\mN=2$ supersymmetric gauge theory of $SU(2)$ with $4$ flavours (see \sref{onesonins} for details), gives the Hilbert series for the coherent component of the one $D_4 = SO(8)$ instanton moduli space (omitting the trivial component $\BC^2$) :
\beq
 \HS_{D_4} (t; x_1, x_2, x_3, x_4) =  \sum_{k=0}^\infty [0,k,0,0]_{D_4} t^{2k} ,
\eeq
Setting these fugacities $y_i$ to 1, we get the unrefined Hilbert series:
\bea
 \HS_{D_4} (t) &=&  \sum_{k=0}^\infty dim[0,k,0,0]_{D_4} t^{2k}  \nn \\
 &=& \frac{(1+t^2)(1+17 t^2+48 t^4+17 t^6+t^8)}{(1-t^2)^{10}} \nn\\
&=& 1 + 28 t^2 + 300 t^4 + \ldots~. 
\eea
An explicit expression for the dimension of each such representation is given by
\bea\label{dimso8} 
\dim~[0,k,0,0]_{D_4}=\frac{(k+1)(k+2)^3(k+3)^3(k+4)(2k+5)}{4320}~.
\eea
Notice that summing the series we get a closed formula with a pole of order $10$ at $t=1$. This means that the space is $10$-complex dimensional, and is in agreement with the fact that the non-trivial component of the one-instanton moduli space for $D_4$ has quaternionic dimension $5$ (the dual Coxeter number $h_{D_4} = 6$).

In general, summing up the unrefined Hilbert series for any group $G$ gives rational functions of the form
\beq \HS_G(t) = \frac{P_G (t^2)}{(1-t^2)^{2h-2}}~, \label{generalHS}\eeq
where $P_G(x)$ is a palindromic polinomial of degree $h_G-1$.


\paragraph{A dimension formula for $Adj^k$.}
Formula (\ref{dimso8}) can be generalised to any classical and exceptional group. Defining
\bea 
G_{a,b}( h, k) = \frac{\binom{(1+a) h/2 - b - 1 + k}{k}}{\binom{(1 - a) h/2 + b - 1 + k}{k}}~,
\eea
the dimension of the $Adj^k$ representation is given by
\bea \label{magicformula} 
\dim~Adj^k = G_{1,1}(h,k) G_{a,b}(h,k) G_{1-a, 1-b}(h,k) \frac{2k + h - 1}{h - 1}~. \eea
where $(a,b,h)$ are given in Table \ref{convtable}.
\footnote{Formula \eref{magicformula} generalises the Proposition 1.1 of \cite{Landsberg01}
\bea
\dim Adj^k = \frac{3c+2k+5}{3c+5} \frac{\binom{k+2c+3}{k}\binom{k+5c/2+3}{k}\binom{k+3c+4}{k}}{\binom{k+c/2+1}{k} \binom{k+c+1}{k}}~, \nn
\eea 
which gives the results for $A_1$, $A_2$, $G_2$, $D_4$, $F_4$, $E_6$, $E_7$ and $E_8$ if we use $c = \frac{1}{3}h_G - 2$.}

\begin{table}
\begin{tabular}{|c|c|c|c||c|}\hline
Lie group & Dynkin label & Dual coxeter  & $(a,b)$ & $\mN=2$ gauge theory\\ 
& of $Adj^k$ & number & & \\
\hline\hline
$A_n = SU(n+1) $  & $ [k,0,\ldots,0,k]$ & $n+1$ & $(1,1)$ & Quiver diagram \ref{f:N2UkUN} \\ \hline
$B_{n \geq 3} = SO(2n+1)$  & $ [0,k,0,\ldots,0]$ & $2n-1$ & $(1,2)$ & Quiver diagram \ref{f:N2SpkSON}  \\ \hline
$C_{n \geq 2} = Sp(2n)$  & $ [2k,0,\ldots,0]$ & $n+1$ & $(1,1/2)$ & Quiver diagram \ref{f:N2SpN} \\ \hline
$D_{n \geq 4} = SO(2n)$  & $ [0,k,0,\ldots,0]$ & $2n-2$ & $(1,2)$  & Quiver diagram \ref{f:N2SpkSON} \\ \hline\hline
$ E_{6}$  & $ [0,k,0,0,0,0]$ & $12$ & $(1/3,0)$ & $3$ M5s on 3-punctured sphere\\ \hline
$ E_{7,8}$  & $ [k,0,\ldots,0]$ & $18,30$ & $(1/3,0)$ & $4,6$ M5s on 3-punctured sphere\\ \hline
$ F_4$  & $ [k,0,0,0]$ & $9$ & $(1/3,0)$ & \\ \hline
$ G_2$  & $ [0,k]$ & $4$ & $(1/3,0)$ & \\ \hline
\end{tabular}\label{convtable}
\caption{Useful information on classical and exceptional groups.  The last column indicates the $\cN=2$ gauge theories, for which the Higgs branch is identified with the corresponding moduli space of instantons on $\BR^4$.}
\end{table}

\section{Gauge theories on D$p$-D$(p+4)$ brane systems}\label{classical}
The moduli space of instantons is known to be the Higgs branch of certain supersymmetric gauge theories \cite{Witten:1995gx, Douglas:1995bn, Douglas:1996uz}. For classical gauge groups there is an explicit construction, while for exceptional gauge groups there is a puzzle on how to explicitly write it down.
Below we recall the string theory embedding of the gauge theories for classical gauge groups as worldvolume theories of D$p$ branes in backgrounds of D$(p+4)$ branes and summarize the gauge theory data for these theories.  It is perhaps convenient to take $p=3$, so that the worldvolume theories have $\cN=2$ supersymmetry in $(3+1)$ dimensions. The presence of these branes breaks space-time into $\BR^{1,3}\times \BC^2\times \BC$. There is a $U(2)$ symmetry that acts on the $\BC^2$ and acts as an $R$ symmetry on the different supermultiplets in the theory. This symmetry is used below to distinguish some of the gauge invariant operators.

The gauge theory on the D3 branes is most conveniently written in terms of $\cN = 2$ quiver diagrams but for the purpose of computing the Hilbert series, it is more convenient to work using an $\cN = 1$ notation. Section \ref{quiver} summarizes the basic rules of translating an $\cN=2$ quiver diagram to an $\cN=1$ quiver diagram with a superpotential.

\subsection{Quiver diagrams}
\label{quiver}
 
\begin{figure}[ht]
\begin{center}
  \vskip-0.5cm
  \includegraphics[totalheight=5cm]{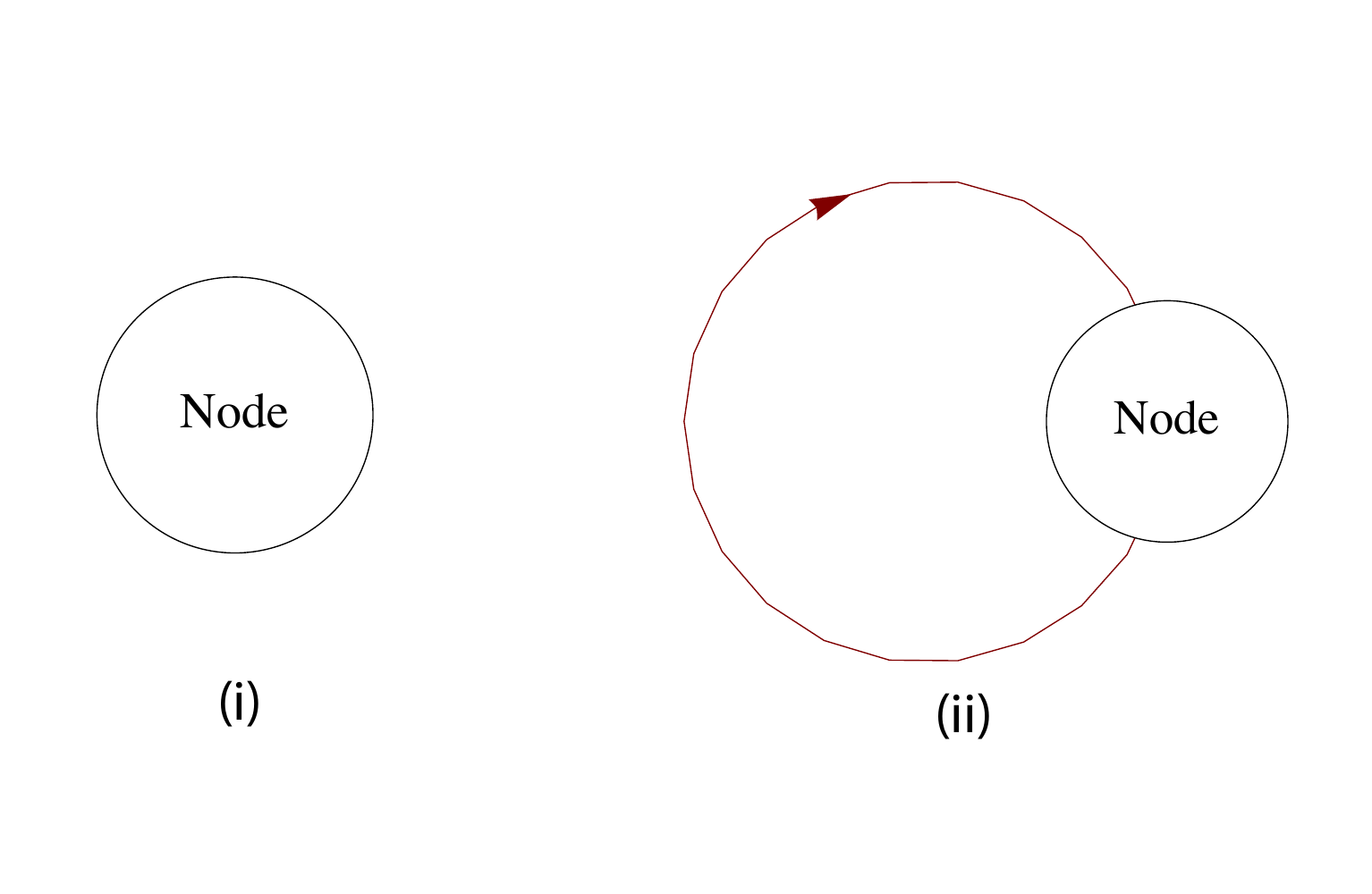}
 \vskip-0.5cm
 \caption{A node in the $\cN=2$ quiver diagram (labelled (i)) becomes a node with an adjoint chiral multiplet in the $\cN=1$ quiver diagram (labelled (ii)).}
  \label{f:cnode}
\end{center}
\end{figure}

 To write down a Lagrangian for a gauge theory with $\cN=2$ supersymmetry it is enough to specify the gauge group, transforming in a vector multiplet, and the matter fields, transforming in hyper multiplets. This can be simply summarized by a quiver with 2 objects - nodes and lines but nevertheless has a two-fold ambiguity on how to assign the objects. A traditional mathematical approach, first introduced to the string theory literature in \cite{Douglas:1996sw}, is to assign nodes to vector multiplets and lines to hyper multiplets. This is the so called quiver diagram used below. The more physically inspired approach \cite{Hanany:1996ie}, is to assign lines to vector multiplets and nodes to hyper multiplets. This notation turns out to be more useful when the hyper multiplets carry more than two charges. On the other hand, to write down the Lagrangian for a gauge theory with $\cN=1$ supersymmetry the data which is needed consists of 3 objects: the gauge group, the matter fields, and the interaction terms written in the form of a superpotential. This can be summarized by an oriented quiver, namely it has arrows which are absent in the $\cN=2$ quiver, and is supplemented by a superpotential $W$. A simple dictionary exists between the two formulations.
It goes as follows:
\bi
\item A node in the $\cN=2$ quiver diagram becomes a node with an adjoint chiral multiplet in the $\cN=1$ quiver diagram. This adjoint chiral multiplet comes from the $\cN=2$ vector multiplet which decomposes as a $\cN=1$ vector multiplet and a $\cN=1$ chiral multiplet.  The map is shown in \fref{f:cnode}.

\begin{figure}[ht]
\begin{center}
  \vskip-3cm
  \includegraphics[totalheight=8.5cm]{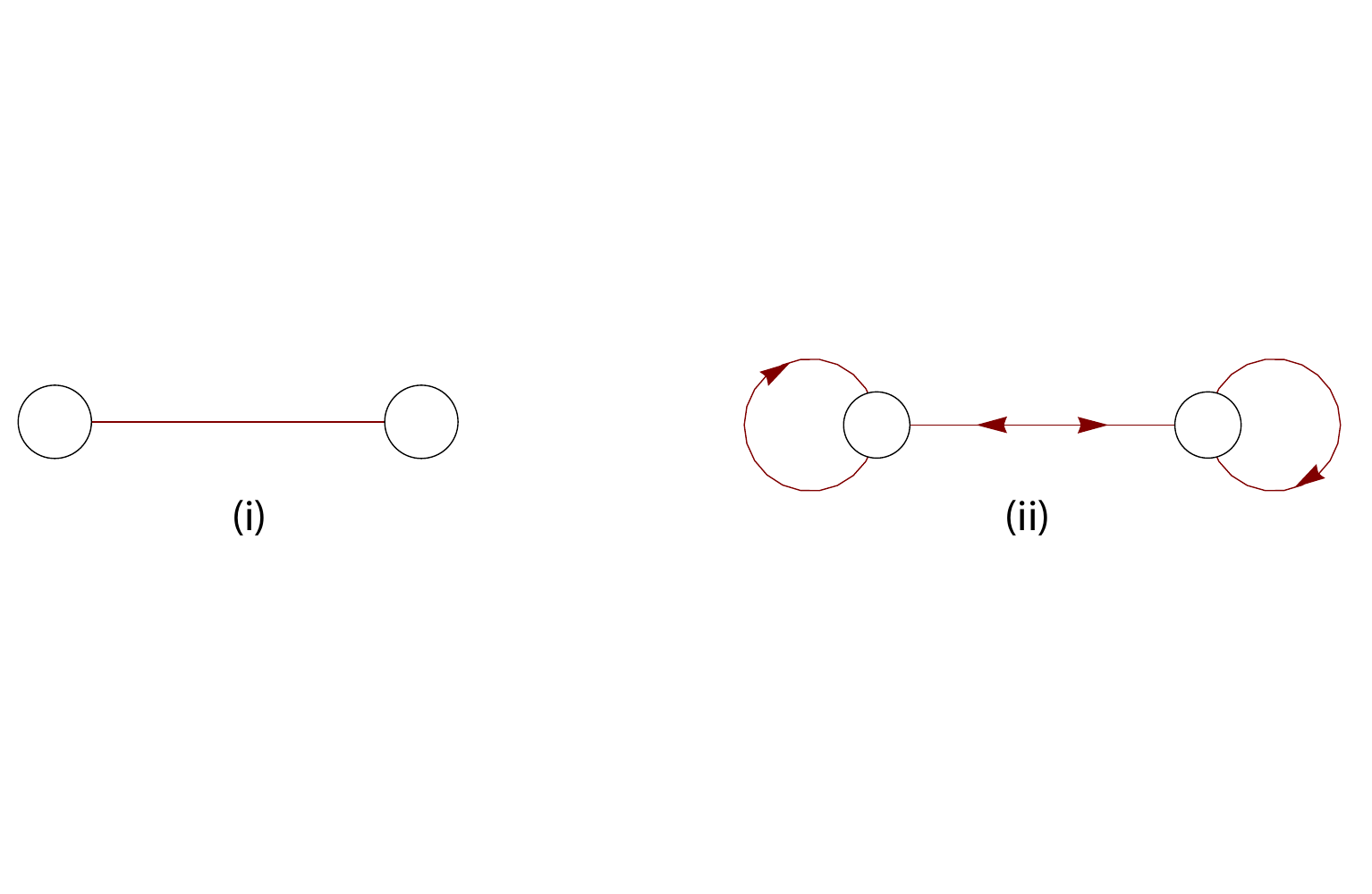}
  \vskip-3cm
 \caption{A line in the $\cN=2$ quiver diagram (labelled (i)) becomes a bi-directional line in the $\cN=1$ quiver diagram (labelled (ii)).}
  \label{f:cline}
\end{center}
\end{figure}

\item A line in the $\cN=2$ quiver diagram becomes a bi-directional line in the $\cN=1$ quiver diagram. This is shown in
\fref{f:cline}.

\begin{figure}[ht]
\begin{center}
  \includegraphics[totalheight=2.8cm]{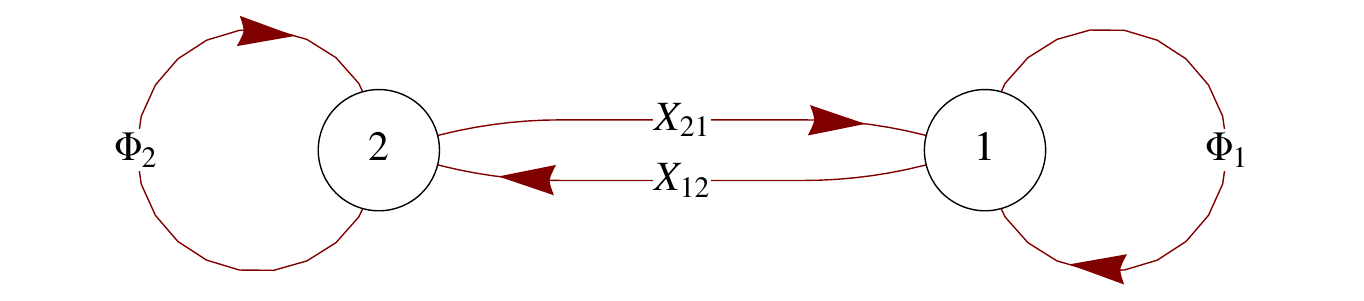}
 \caption{An $\cN=1$ quiver diagram with the superpotential 
 : $X_{21} \cdot \Phi_1 \cdot X_{12} - X_{12} \cdot \Phi_2 \cdot X_{21}$. }
  \label{f:sup}
\end{center}
\end{figure}

\item The superpotential is given by the sum of contributions from all lines in the $\cN=2$ quiver diagram. Each line stretched between two nodes in the $\cN=2$ quiver diagram contributes two cubic superpotential terms. Let the two nodes be labeled by 1 and 2.  Associated with each node, there is an adjoint field denoted respectively by $\Phi_1$ and $\Phi_2$.  A line connecting between two nodes contains two $\cN=1$ bi-fundamental chiral multiplets $X_{12}$ and $X_{21}$.  (The $\cN=1$ quiver diagram is drawn in \fref{f:sup}.)  The corresponding superpotential term is written as an adjoint valued mass term for the $X$ fields:
\bea
X_{21} \cdot \Phi_1 \cdot X_{12} - X_{12} \cdot \Phi_2 \cdot X_{21}~, \label{supsc}
\eea
This notation means as follows. Denote the rank of nodes 1 and 2 by $r_1$ and $r_2$ respectively. then $\Phi_1, \Phi_2, X_{12}, X_{21} $ can be chosen to be $r_1 \times r_1, r_2\times r_2, r_1\times r_2, r_2\times r_1$ matrices, respectively. The $\cdot$ corresponds to matrix multiplication and an impiicit trace is assumed.
Note that this is a schematic notation which does not specify the index contraction whose details depend on the gauge and flavour groups. As a special case, a line from one node to itself would naturally produce a commutator.

\ei

\begin{figure}[ht]
\begin{center}
  \includegraphics[totalheight=3.5cm]{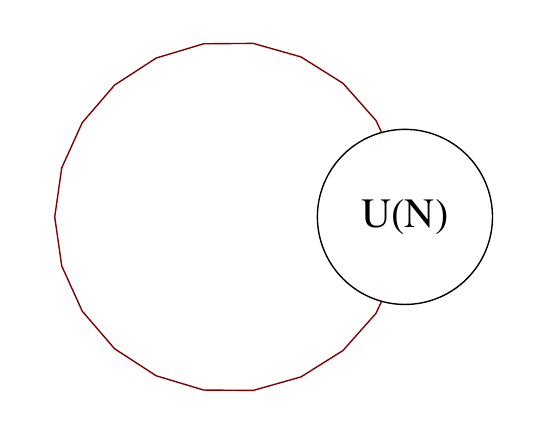}
 \caption{The $\cN=2$ quiver diagram for the $\cN=4$ SYM theory with gauge group $U(N)$.  The loop around the $U(N)$ gauge group denotes an adjoint hypermultiplet.}
  \label{f:N2sym}
\end{center}
\end{figure}

\begin{figure}[ht]
\begin{center}
\vskip-0.5cm
  \includegraphics[totalheight=6cm]{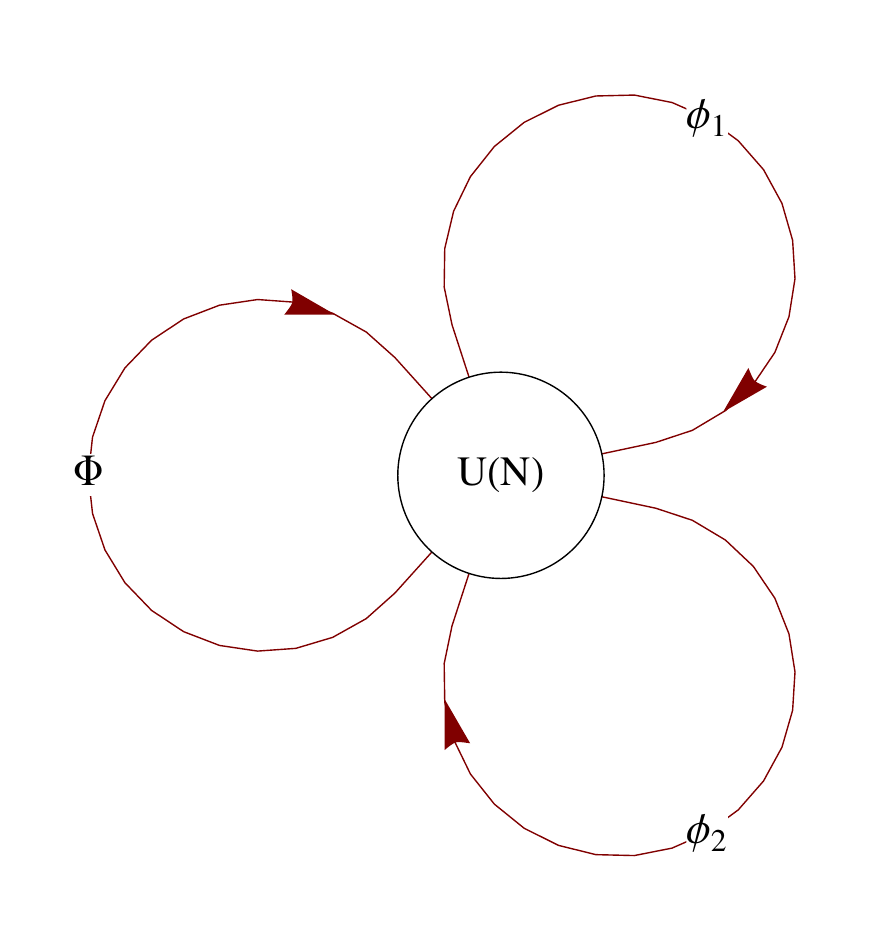}
 \caption{The $\cN=1$ quiver diagram of the $\cN=4$ SYM theory.  The adjoint field $\Phi$ comes from the $\cN=2$ vector multiplet, whereas the adjoint fields $\phi_1, \phi_2$ come from the $\cN=2$ adjoint hypermultiplet.  The superpotential is $W = \tr (\phi_1 \cdot \Phi \cdot \phi_2 - \phi_2 \cdot \Phi \cdot \phi_1) = \tr \left( \Phi \cdot [ \phi_1,  \phi_2] \right)$.}
  \label{f:N1sym}
\end{center}
\end{figure}

As an example, we give the $\cN=2$ and $\cN=1$ quiver diagrams for the $U(N)$ $\cN=4$ super Yang-Mills (SYM) respectively in \fref{f:N2sym} and \fref{f:N1sym}.

\subsection{$k$ $SU(N)$ instantons on $\BC^2$}

\begin{figure}[ht]
\begin{center}
  \includegraphics[totalheight=3cm]{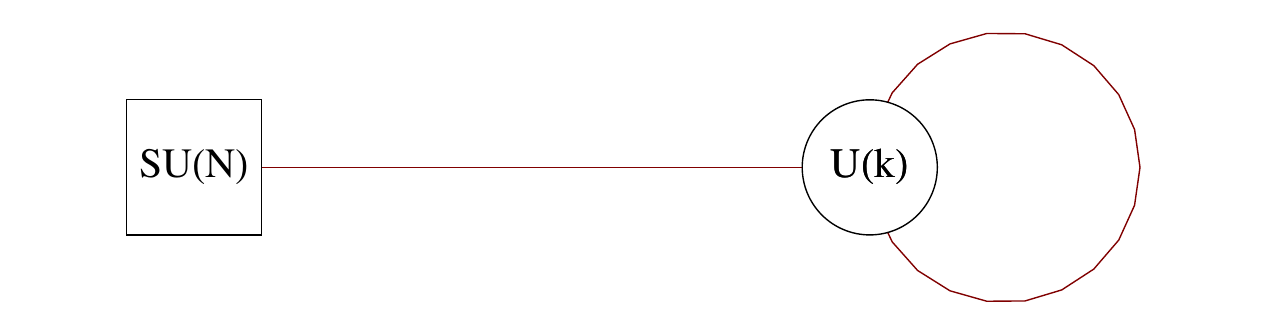}
 \caption{The $\cN=2$ quiver diagram for $k$ $SU(N)$ instantons on $\BC^2$.  The circular node represents the $U(k)$ gauge symmetry and the square node represents the $SU(N)$ flavour symmetry. The line connecting the $SU(N)$ and $U(k)$ groups denotes $kN$ bi-fundamental hypermultiplets, and the loop around the $U(k)$ group denotes the adjoint hypermultiplet.}
  \label{f:N2UkUN}
\end{center}
\end{figure}

With this quiver notation it is now very simple to write down the gauge theory living on the world volume of $k$ D3 branes in the background of $N$ D7 branes. In fact, the brane system very naturally forms a quiver and we can just write down a dictionary between the branes and the objects in the quiver. We will write down the theory using $\cN=2$ quivers and then translate it to $\cN=1$ quivers. First, the gauge theory on $k$ D3 branes is the well known $\cN=4$ supersymmetric theory with gauge group $U(k)$ depicted in \fref{f:N2sym}. The D7 branes are heavier and therefore give rise to a global $U(N)$ symmetry on the worldvolume of the D3 branes.  As discussed below, the global $U(1)$ of $U(N)$ may be absorbed into the local $U(1)$ of $U(k)$; therefore global $SU(N)$ symmetry is represented by a square node with index $N$.  Finally strings stretched between the D3 branes and the D7 branes are represented by a line connecting the circular node to the square node. The resulting quiver is depicted in \fref{f:N2UkUN}.

\begin{figure}[ht]
\begin{center}
  \includegraphics[totalheight=6cm]{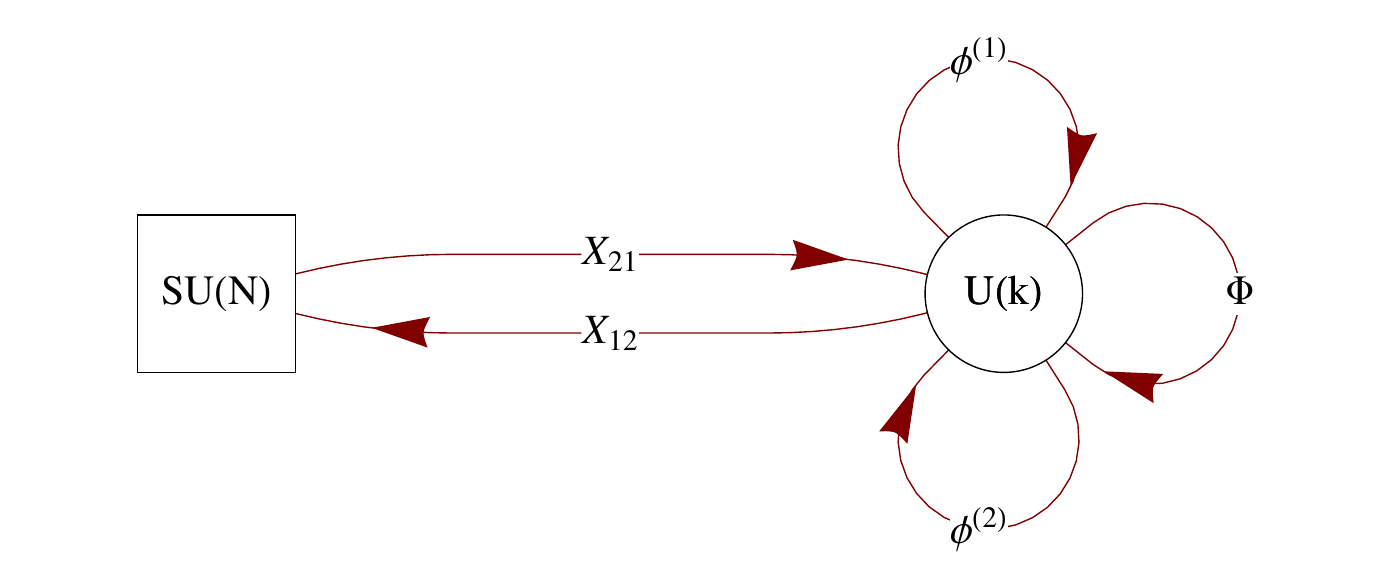}
 \caption{Flower quiver; The $\cN=1$ quiver diagram for $k$ $SU(N)$ instantons on $\BC^2$ with the corresponding superpotential, $W =X_{21} \cdot \Phi \cdot X_{12} + \epsilon_{\alpha \beta} \phi^{(\alpha)} \cdot \Phi \cdot \phi^{(\beta)}$.}
  \label{f:N1UkUN}
\end{center}
\end{figure}

It is now straightforward to apply the rules of \sref{quiver} to write down the $\mN=1$ quiver diagram which is depicted in \fref{f:N1UkUN} and its corresponding superpotential. To write down the superpotential we need explicit notation for the quiver fields and the line between the circular node and the square node corresponds to two chiral fields denoted by $Q$ and $\tQ$. Putting this together, $W$ takes the form
\bea
W &=& X_{21} \cdot \Phi \cdot X_{12}+ \left( \phi^{(1)} \cdot \Phi \cdot \phi^{(2)} -  \phi^{(2)} \cdot \Phi \cdot \phi^{(1)} \right) \nn \\
&=& X_{21} \cdot \Phi \cdot X_{12} + \epsilon_{\alpha \beta} \phi^{(\alpha)} \cdot \Phi \cdot \phi^{(\beta)}~ .\label{WUk}
\eea
Note that the rules for writing the quiver imply the existence of another term coming from the adjoint in the vector multiplet of the D7 branes. This term corresponds to an adjoint $U(N)$ valued mass term for the bifundamental fields $X_{12}, X_{21}$. In this paper we will not treat this mass term and set it to 0, even though it is interesting to consider the effects of such a term. The adjoint fields are parametrizing the position of the D3 branes in $\BC^2$. Since there is a natural $U(2)_g = SU(2)_g\times U(1)_g$ symmetry that acts on $\BC^2$, the fields $\phi_1$ and $\phi_2$ transform as a doublet of $SU(2)_g$ symmetry and with charge 1 under $U(1)_g$. The superpotential should therefore be invariant under $SU(2)_g$ and carry charge $2$ under $U(1)_g$. 

We list the charges and the representations under which the fields transform in \tref{t:chargeU}.
\begin{table}[htdp]
\begin{center}
\begin{tabular}{|c|c|c|c|c|c|c|}
\hline
Field & \multicolumn{2}{|c|}{$U(k)$} & \multicolumn{2}{|c|}{$U(N)$}& $SU(2)_g$ & $U(1)_g$\\
\hline
& $SU(k)$ & $U(1)$ &  $SU(N)$ & $U(1)$ & global & global \\
\hline
Fugacity: & $z_1, \ldots, z_{k-1}$ & $z$ & $x_1, \ldots, x_{N-1}$  & $q$ & $x$ &  $t$ \\
\hline
$\Phi$ & $[1,0,\ldots,0,1]$ & $0$ &  $[0,\ldots,0]$ & $0$ & [0] &  0\\
$\phi^{(1)}, \phi^{(2)}$ & $[1,0,\ldots,0,1]$ & $0$ &  $[0,\ldots,0]$ & $0$ & [1]  & 1 \\
$X_{12}$ & $[1,0, \ldots, 0]$ & $1$ & $[0, \ldots, 0,1]$ & $-1$ & [0] & 1\\ 
$X_{21}$ & $[0,0, \ldots,0, 1]$ & $-1$ & $[1,0 \ldots, 0]$ & $1$ & [0] & 1\\ 
$\tr \Phi$ & [0,\ldots,0] & 0 & [0,\ldots,0] & 0 & [0] & 0 \\
$\tr \phi^{(1)}, \tr \phi^{(2)}$ & [0,\ldots,0] & 0 & [0,\ldots,0] & 0 & [1] & 1 \\
\hline
\end{tabular}
\caption{The charges and the representations under which various fields transform.  The fugacites of each field are assigned according to this table.  The $U(2)_g$ global symmetry acts on $\phi^{(1)}$ and $\phi^{(2)}$. It is the symmetry group of $\BC^2$, the trivial component in the moduli space.}
\label{t:chargeU}
\end{center}
\end{table}%

From \tref{t:chargeU}, it can be seen that the $U(1)$ of $U(N)$ can be absorbed into the local $U(1)$ (\emph{e.g.} by means of redefining the fugacity $z/q$).  From the brane perspective, the vector multiplet of the local $U(1)$ contains a scalar which parametrises the position of the D3-brane in the 
directions transverse to the D7 branes. 
One can set the origin of these directions to be at the CoM of the D7-branes and thereby eliminate the corresponding  background $U(1)$ vector multiplet.

Let us compute the quaternionic dimension of the Higgs branch.  From the $\cN=2$ quiver diagram, the line connecting the $SU(N)$ and $U(k)$ groups denotes $kN$ hypermultiplets, and the loop around the $U(k)$ group denotes $k^2$ hypermultiplets.  Hence, we have in total $kN+k^2$ quarternionic degrees of freedom.   On a generic point on the Higgs branch, the gauge group $U(k)$ is completely broken and hence there are $k^2$ broken generators.  As a result of the Higgs mechanism, the vector multiplet gains $k^2$ degrees of freedom and becomes massive.  Hence, the $(kN+k^2)-k^2 = kN$ quarternionic degrees of freedom are left massless.  Thus, the Higgs branch is $kN$ quaternionic dimensional or $2kN$ complex dimensional:
\bea
\dim_{\BC} \MH_{k,N} = 2kN = 2kh~. \label{UkNdim}
\eea
This agrees with the dual coxeter number of $SU(N)$ which is $h_{SU(N)}=N$.

From the brane perspective, the VEV of the scalar $\Phi$ correspond to the position of the D3-branes along the 
directions transverse to the D7-branes.  On the Higgs branch, the gauge fields become massive freezing the whole vector multiplet and hence $\langle \Phi \rangle = 0$, setting the D3 branes to lie within the D7 branes and possibly form bound states. The hypermultiplets acquire non-zero VEVs at a generic point on the Higgs branch that parametrize all possible bound states of D3 and D7 branes. 
From the point of view of the D7 brane gauge theory, the D3 branes are interpreted as instantons and hence, the moduli space of classical instantons on $\BC^2$ is identified with the Higgs branch of the quiver theory \cite{Douglas:1995bn}.

\subsubsection{One $SU(N)$ instanton: $k=1$}
The gauge theory for 1 $SU(N)$ instanton on $\BC^2$ is particularly simple and lives on the world volume of 1 D3 brane, $k=1$. The gauge group is $U(1)$ and the adjoints $\Phi, \phi_1, \phi_2$ are simply complex numbers, and hence the second term of \eref{WUk} vanishes,

\beq
W =  X_{21} \cdot \Phi \cdot X_{12} .
\eeq


\paragraph{The Higgs branch.} On the Higgs branch, $\Phi = 0$ and $X_{12} \cdot  X_{21}=  0$.  The space of F-term solutions (which we will call the F-flat space and denote by $\f$) is obviously a complete intersection.  Using \eref{UkNdim} the dimension of the moduli space is $2N$.  On the other hand there are $2N$ bifundamental fields $X_{12}, X_{21}$ and 2 $\phi$'s which are subject to 1 relation. This gives an F-flat moduli space which is $2N+1$ dimensional and after imposing the D-term equations we get a $2N$ dimensional moduli space, as expected. The F-flat Hilbert series can be written down according to \tref{t:charge} as\footnote{The {\bf plethystic exponential (PE)} of a multi-variable function $g (t_1 , \ldots , t_n )$ that vanishes at the origin, $g (0,\ldots, 0) = 0$, is defined to be $\PE [g (t_1 , \ldots , t_n )] := \exp \left ( \sum_{r=1}^\infty \frac{g(t_1^r,\ldots,t_n^r)}{r}\right )$. The reader is referred to \cite{pleth1, pleth2, apercu} for more details.}
\bea
\gf_{k=1,N}(t,x_1, \ldots, x_{N-1},x, q, z)  &=& (1-t^2) \PE \Big[ [1]_{SU(2)_g} t + [1,0, \ldots, 0]_{SU(N)} \frac{t z}{q}  \nn \\
&& +  [0,0, \ldots, 0, 1]_{SU(N)} \frac{t q}{ z} \Big]~. \label{fffull}
\eea
Note that the first term in the square bracket corresponds to $\phi^{(1)}$ and $\phi^{(2)}$, the second term corresponds to $X_{12}$ and the third term correspond to $X_{21}$, and the factor in front of the $\PE$ corresponds to the relation.

Notice from \eref{fffull} that the $U(1)$ of $U(N)$ can in fact be absorbed into the local $U(1)$. This can be seen by redefining the fugacity for the local $U(1)$ as
\bea
w = \frac{z}{q}~,
\eea
and rewrite
\bea
\gf_{k=1,N}(t,x_1, \ldots, x_{N-1},x, w)  &=& (1-t^2) \PE \Big[ [1]_{SU(2)_g} t  + [1,0, \ldots, 0]_{SU(N)} t w  \nn \\
&& +  [0,0, \ldots, 0, 1]_{SU(N)} \frac{t}{w} \Big]~. 
\eea
The right hand side can explicitly be written as a rational function:
\bea
&& (1-t^2)  \times \frac{1}{(1- t x)(1-\frac{t}{x})} \times \frac{1}{ \left(1-t  w  x_1\right) \left(1- \frac{t w}{ x_{N-1}} \right) \prod_{k=2}^{N-1} (1-t  w  \frac{x_k}{x_{k-1}} ) } \nn \\
&& \times \frac{1}{ \left(1-\frac{t}{ w}  \frac{1}{x_1}\right) \left(1- \frac{t}{w}  x_{N-1} \right) \prod_{k=2}^{N-1} (1- \frac{t}{ w}  \frac{x_{k-1}}{x_{k}} ) }~.\label{pe}
\eea

\paragraph{The Hilbert series.} Now we project \eref{pe} onto the gauge invariant subrepresentation by performing an integration over the $U(1)$ gauge group\footnote{This is called the {\bf Molien-Weyl integral formula} (see, \emph{e.g.}, \cite{pleth2, apercu}).}.  The Hilbert series of the Higgs branch is therefore given by
\bea
\gH_{k=1, N}(t,x_1, \ldots, x_{N-1},x) = \frac{1}{2 \pi i} \oint_{|w| =1} \frac{\ud w}{w}  \gf_{k=1,N}(t,x_1, \ldots, x_{N-1},x,w)~. \label{molien}
\eea
Using the residue theorem on \eref{pe}, where the poles are located at\footnote{Note that $|t| < 1$ and only poles located inside the unit circle $|w|=1$ are included.}
\bea
w = t \frac{1}{x_1},~t \frac{x_1}{x_2},~\ldots,~ t \frac{x_{N-2}}{x_{N-1}},~t x_{N-1}~,
\eea 
we can write the Hilbert series in terms of representations as
\bea
\gH_{k=1, N}(t,x_1, \ldots, x_{N-1},x) &=&  \frac{1}{(1- t x)\left(1-\frac{t}{x} \right)} \sum_{k=0}^\infty [k, 0, \ldots, 0,k]_{SU(N)} t^{2k} .
\eea
The factor $\frac{1}{(1- t x)(1-\frac{t}{x})}$  indicates the Hilbert series for the complex plane $\BC^2$, whose symmetry is $U(2)_g$ (with the fugacities $t, x$).  This space $\BC^2$ is parametrised by $\phi^{(1)}$ and $\phi^{(2)}$ and corresponds to the position of the D3-brane inside the D7-branes.  The second factor corresponds to the coherent component of the one $SU(N)$ instanton moduli space.
Unrefining by setting $x_1 = \ldots = x_{N-1} = x =1$, we obtain
\bea
\gH_{k=1, N}(t, 1, \ldots, 1) = \frac{1}{(1-t)^2} \times \frac{\sum_{k=0}^{N-1} \binom{N-1}{k}^2 t^{2k}}{(1-t^2)^{2(N-1)}}~. \label{unrefun}
\eea
The order of the pole $t=1$ is $2N$, and hence the dimension of the Higgs branch is $2N$, in accordance with \eref{UkNdim}. 
Note that \eref{unrefun} can also be derived directly from \eref{molien} as follows.  Setting $x_1 = \ldots = x_{N-1} = x =1$ in \eref{molien}, we obtain
\bea
\gH_{k=1, N}(t, 1, \ldots, 1) = \frac{(1 - t^2)}{(1 - t)^2}  \frac{1}{2 \pi i} \oint_{|w| =1} \frac{\ud w}{w} \frac{1}{(1-t w)^N (1-\frac{t}{w})^N}~.
\eea
The contribution to the integral comes from the pole at $w = t$, which is of order $N$.  Using the residue theorem, we find that
\bea
\gH_{k=1, N}(t, 1, \ldots, 1) = \frac{(1 - t^2)}{(1 - t)^2 } \times \frac{1}{(N-1)!} \frac{\ud^{N-1}}{\ud w^{N-1}} \left[ \frac{w^{N-1}}{(1-t w)^N} \right]_{w =t} 
\eea
Using Leibniz's rule for differentiation, we thus arrive at \eref{unrefun}.

The plethystic logarithm can be written as
\bea
\PL [\gH_{k=1, N}(t,x_1, \ldots, x_{N-1},x)] &=& [1]_{SU(2)_g} t +  [1,0,\ldots,0,1]_{SU(N)} t^2 - \left ([0,1,0, \ldots, 0,1,0] \right . + \nn \\
&& \left . [1, 0, \ldots, 0, 1]+[0, \ldots, 0] \right )_{SU(N)} t^4 + \ldots~.
\eea
Hence, the generators are $\tr \phi^{(1)}, \tr \phi^{(2)}$ at order $t$ and the adjoints [1,0,\ldots,0,1] of $SU(N)$ at the order $t^2$.  The basic relations transform in the $SU(N)$ representation $[0,1,0, \ldots, 0,1,0]+ [1, 0, \ldots, 0, 1]+[0, \ldots, 0]$.

\subsection{$k$ $SO(N)$ instantons on $\BC^2$ }
\begin{figure}[ht]
\begin{center}
  \includegraphics[totalheight=3cm]{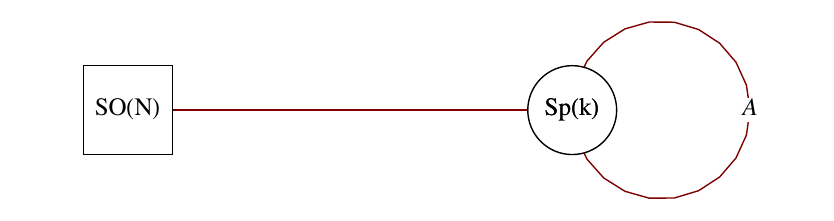}
 \caption{The $\cN=2$ quiver diagram for $k$ $SO(N)$ instantons on $\BC^2$.  The circular node represents the $Sp(k)$ gauge symmetry and the square node represents the $SO(N)$ flavour symmetry. The line connecting the $SO(N)$ and $Sp(k)$ groups denotes $2kN$ half-hypermultiplets, and the loop around the $Sp(k)$ gauge group denotes a hypermultiplet transforming in the (reducible) second rank antisymmetric tensor.}
  \label{f:N2SpkSON}
\end{center}
\end{figure}

As pointed out in \cite{Witten:1995gx}, the moduli space of $k$ $SO(N)$ instantons can be realised on a system of $k$ D$3$-branes with $N$ half D$7$-branes on top of an O7$^-$ orientifold plane. (If the number of branes is odd, the combination of half D7 brane stuck on the O7$^-$ plane form an orientifold plane which is called $\widetilde{O7}^-$ plane.) The brane picture is similar to the one described in the previous subsection and therefore the quiver looks the same. We only need to figure out the action of the orientifold plane on the different objects in the quiver. All together, there are 4 objects in \fref{f:N2UkUN}.
\begin{itemize}
\item The gauge group on the D7 branes is projected to $SO(N)$. This is a global symmetry for the gauge theory on the D3 branes. $\cN=2$ supersymmetry restricts the gauge theory on the D3 branes to be $Sp(k)$. Hence,
\item The gauge group on the D3 branes is projected down to $Sp(k)$.
\item The bi-fundamental fields become bi-fundamentals of $SO(N)\times Sp(k)$.
\item The loop around the $U(k)$ gauge group undergoes a $\BZ_2$ projection which leaves two options - the second rank symmetric or antisymmetric representation of $Sp(k)$. To find which one, we notice that only the anti-symmetric representation is reducible into a singlet plus the rest. Since  the center of mass of the instanton is physically decoupled from the rest of the moduli space, we conclude that the projection is to the antisymmetric representation.
\end{itemize}
The resulting $\mN=2$ quiver diagram is depicted in \fref{f:N2SpkSON}.

\begin{figure}[ht]
\begin{center}
  \includegraphics[totalheight=6cm]{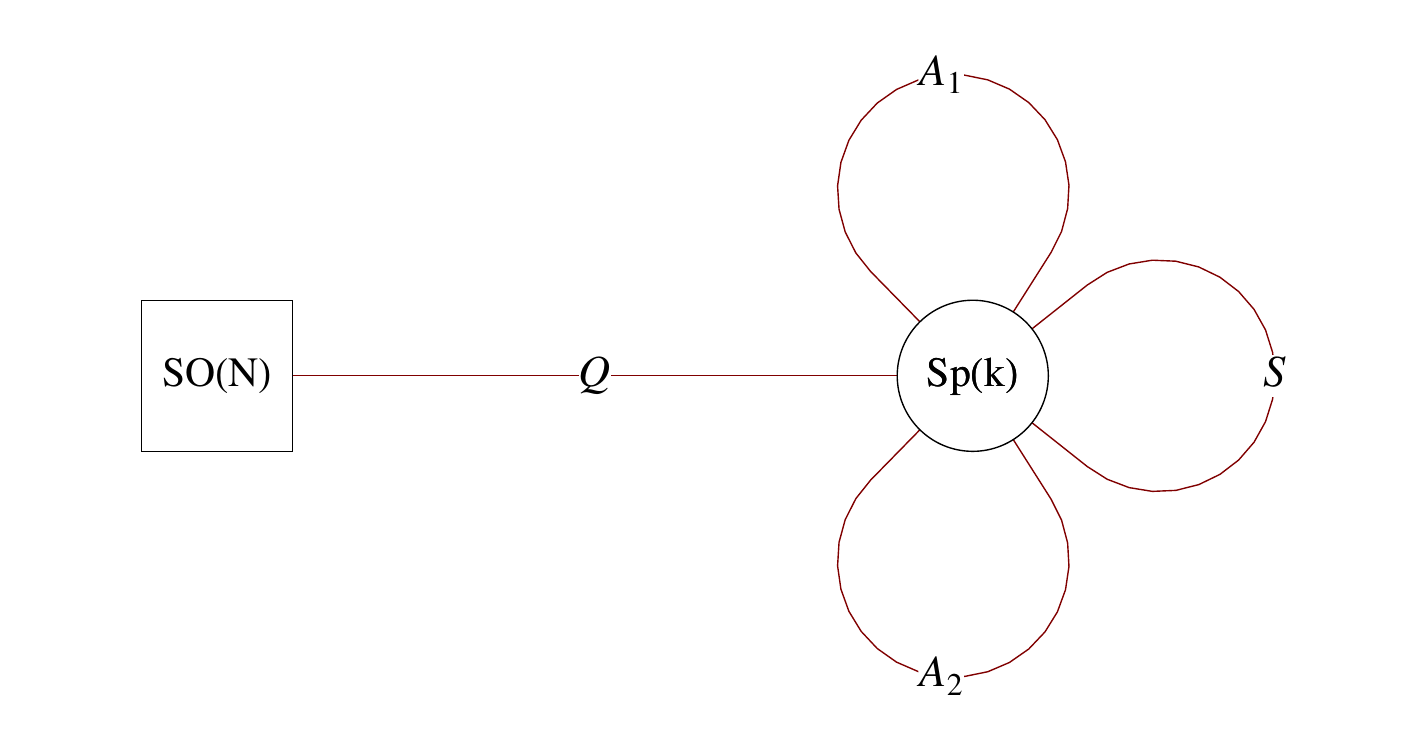}
 \caption{The $\cN=1$ quiver diagram for $k$ $SO(N)$ instantons on $\BC^2$. The chiral multiplet transforming in the second rank symmetric tensor (adjoint field) of $Sp(k)$ is denoted by $S$ and the second rank antisymmetric tensors are denoted by $A_1, A_2$.  The superpotential is given by $W = Q \cdot S \cdot Q + \epsilon_{\alpha \beta} A_\alpha \cdot S \cdot A_\beta $.}
  \label{f:N1SpkSON}
\end{center}
\end{figure}

Using the rules of \sref{quiver} it is easy to find the $\mN=1$ quiver diagram given in \fref{f:N1SpkSON} and the superpotential,
\bea
W &=& Q \cdot S \cdot Q + \left( A_1 \cdot S \cdot A_2 - A_2 \cdot S \cdot A_1 \right) \nn \\
&=& Q \cdot S \cdot Q + \epsilon_{\alpha \beta} A_\alpha \cdot S \cdot A_\beta~,
\label{WSpk}
\eea
where we have suppressed the contractions over the gauge indices by the tensor $J^{ab}$ (an invariant tensor of $Sp(k)$) and the contractions over the flavour indices by $\delta_{ij}$ (an invariant tensor of $SO(N)$).  The epsilon tensor $\epsilon_{\alpha \beta}$ in the second line is an invariant tensor of the global 
$SU(2)$ symmetry which interchanges $A_1$ and $A_2$. The mass term for $Q$ coming from the adjoint of $SO(N)$ is set to 0.

Let us compute the quaternionic dimension of the Higgs branch.  From the $\cN=2$ quiver diagram, the lines connecting the $SO(N)$ and $Sp(k)$ groups denotes $2kN$ half-hypermultiplets (equivalently, $kN$ hypermultiplets), and the loop around the $Sp(k)$ group gives $k(2k-1)$ hypermultiplets.  Hence, we have in total $kN+k(2k-1)$ quarternionic degrees of freedom.   On the Higgs branch, $Sp(k)$ is completely broken and hence there are $k(2k+1)$ broken generators.  As a result of the Higgs mechanism, the vector multiplet gains $k(2k+1)$ degrees of freedom and becomes massive.
Hence, the $kN+k(2k-1)-k(2k+1) = k(N-2)$ degrees of freedom are left massless.  Thus, the Higgs branch is $k(N-2)$ quaternionic dimensional or $2k(N-2)$ complex dimensional:
\bea
\dim_{\BC} \MH_{k,N} = 2k(N-2) = 2kh_{SO(N)} ~. \label{SpkNdim}
\eea
Note that $h_{SO(N)} = N-2$ is the dual coxeter number of the $SO(N)$ group.

The charges and the representations under which the fields transform are given in \tref{t:charge} \cite{Seiberg:1994aj}.
\begin{table}[htdp]
\begin{center}
\begin{tabular}{|c|c|c|c|c|c|c|}
\hline
Field & $Sp(k)$ &$SO(N)$& $SU(2)_g$ &  $U(1)_g$\\
\hline
Fugacity: & $z_1, \ldots, z_{k}$& $x_1, \ldots, x_{\lfloor N/2 \rfloor}$ & $x$ & $t$ \\
\hline
$S$ & $[2,0,\ldots,0]$  &  $[0,\ldots,0]$  & [0] & 0\\
$A_1, A_2$ & $[0,1,0,\ldots,0]+[0, \ldots, 0]$  &  $[0,\ldots,0]$  & [1] &  1 \\
$Q$ & $[1,0, \ldots, 0]$  & $[1,0, \ldots, 0]$  & [0] &  1\\ 
\hline
\end{tabular}
\caption{The charges and the representations under which various fields transform.  The fugacites of each field are assigned according to this table.}
\label{t:charge}
\end{center}
\end{table}%

\subsubsection{One $SO(N)$ instanton on $\BC^2$: $k=1$} \label{onesonins}
\label{oneson}
In the special case $k=1$, the gauge group is $Sp(1)=SU(2)$ and the superpotential \eref{WSpk} becomes
\bea
W_{k=1} = \epsilon^{ab} \epsilon^{cd}  Q^i_a S_{bc} Q^i_d~.
\eea

\paragraph{The Higgs branch.}
The Higgs branch is given by the F-term conditions: $S = 0$ and $Q^i_a Q^i_b + Q^i_b Q^i_a =0$, and the D-term condition. The Hilbert series of the F-flat moduli space is
\bea
\gf(t, z, x_1, \ldots, x_{\lfloor N/2 \rfloor},x) &=& \left(1-t^2\right) \left(1-\frac{t^2}{z^2}\right) \left(1-t^2 z^2\right) \PE \Big[ [1]_{SU(2)_g} t  \nn \\
&& +[1,0,\ldots,0]_{SO(N)} t \left(z+\frac{1}{z} \right) \Big]~. \label{ffso}
\eea
We note that the relation transforms in the representation $[2]$ of $Sp(1)$ and that the F-flat moduli space is a complete intersection of dimension $2+2N-3=2N-1$.  Noting that the characters of the fundamental representations of $B_{n} = SO(2n+1)$ and $D_{n}=SO(2n)$ respectively are
\bea
 \left[1,0, \ldots,0 \right]_{B_n} (x_a) &=& 1+ \sum_{a=1}^{n} \left( x_a + \frac{1}{x_a} \right) ~, \nn \\
 \left[1,0, \ldots,0 \right]_{D_n} (x_a) &=&  \sum_{a=1}^{n} \left( x_a + \frac{1}{x_a} \right)~, \label{charfunds}
\eea
we can write down \eref{ffso} as a rational functional function
\bea
\gf(t, z, x_1, \ldots, x_n,x)_{B_n, D_n} &=& \frac{\left(1-t^2\right)  }{(1-t x)(1-t/x)} \times \nn \\
&& \frac{\left(1-\frac{t^2}{z^2}\right) \left(1-t^2 z^2\right)}{(1-t)^\delta \prod_{a=1}^n (1-t z x_a)(1- \frac{t z}{x_a})(1-\frac{t}{z} x_a)(1- \frac{t }{z x_a})}~, \nn \\
\eea
where $\delta =1$ for $B_n$ and $\delta=0$ for $D_n$.

Performing the Molien-Weyl integral over the gauge group $Sp(1)$, we obtain the Higgs branch Hilbert series as
\bea
\gH(t, x_1, \ldots, x_n, x)_{B_n, D_n} &=& \frac{1}{2\pi i} \oint_{|z|=1} \ud z \left( \frac{1-z^2}{z} \right) \gf(t, z, x_1, \ldots, x_n,x)_{B_n, D_n} \nn \\
&=& \frac{1}{(1-t x)(1-t/x)} \times \sum_{k=0}^\infty [0,k,0, \ldots,0]_{_{B_n, D_n}} t^{2k}~, \label{mwsp}
\eea
where the contributions to the integral come from the poles:
\bea
z = t x_1, \ldots, t x_n, \frac{t}{x_1}, \ldots, \frac{t}{x_n}~.
\eea

The factor $\frac{1}{(1-t x)(1-t/x)}$ is the Hilbert series for $\BC^2$ (whose symmetry is $U(2)_g$) and is parametrised by the singlets in $A_1, A_2$; this corresponds to the position of the D3-brane inside the D7-branes.  The second factor corresponds to the coherent component of the one $SO(N)$ instanton moduli space.

\paragraph{Example: $N=8$.}  
The expression \eref{ffso} can be written as a rational function:
\bea
\frac{\left(1-t^2\right)}{(1-t x)(1-t/x)}  \times \frac{ \left(1-\frac{t^2}{z^2}\right) \left(1-t^2 z^2\right)}{ \prod_{a=1}^4 (1-t z x_a)(1- \frac{t z}{x_a})(1-\frac{t x_a}{z})(1- \frac{t }{z x_a})}~.
\eea
The poles which contribute to the Molien-Weyl integral \eref{mwsp} are
\bea
z = t x_1, \ldots, t x_4, \frac{t}{x_1}, \ldots, \frac{t}{x_4}~.
\eea
The integral \eref{mwsp} gives 
\bea
\gH(t, x_1, \ldots, x_4, x) &=& \frac{1}{(1-t x)(1-t/x)} \times \sum_{k=0}^\infty [0,k,0,0]_{SO(8)} t^{2k}~. \label{gHso8}
\eea
Unrefining by setting $x_1 = \ldots =x_4= x=1$, we obtain
\bea
\gH(t, 1,1,1,1,1) = \frac{1}{(1-t)^2} \times \frac{\left(1+t^2\right) \left(1+17 t^2+48 t^4+17 t^6+t^8\right)}{\left(1-t^2\right)^{10}}~.
\eea
Observe that the pole at $t=1$ is of order $12$, and so the Higgs branch is indeed 12 dimensional, in agreement with \eref{SpkNdim}.
The plethystic logarithm is
\bea
\PL \left[ \gH(t, x_1,x_2,x_3,x_4,x) \right] &=& [1]_{SU(2)_g} t  + [0,1,0,0]_{SO(8)} t^2 - ([2,0,0,0]+[0,0,2,0] \nn \\
&& +[0,0,0,2]+ [0,0,0,0])_{SO(8)} t^4 + \ldots~,
\eea
indicating that the relations are invariant under the triality of $SO(8)$.

\subsection{$k$ $Sp(N)$ instantons on $\BC^2$}
As pointed out in \cite{Douglas:1995bn}, the moduli space of $k$ $Sp(N)$ instantons can be realised on a system of $k$ D$3$-branes with $N$ D$7$-branes on top of an O7$^+$ orientifold plane.  As a result, the gauge group is projected to $SO(k)$, \footnote{For $k=1$ we take the convention that $SO(1)$ is $\BZ_2$. For higher values of $k$, the computations in this paper do not distinguish between a gauge group $O(k)$ and a gauge group $SO(k)$ and hence this $\BZ_2$ ambiguity is ignored. 
}
and the scalar in the vector multiplet becomes an antisymmetric tensor, denoted by $A_{ab}$ (where the $SO(k)$ gauge indices take values $a,b=1,\ldots,k$).  The adjoint hypermultiplet becomes a symmetric tensor, as it is the reducible second rank tensor of $SO(k)$, and is denoted by two chiral multiplets $S_1$ and $S_2$.  Since representations of the $SO(k)$ group are real, the flavour symmetry is $Sp(N)$ and we have $2kN$ half-hypermultiplets.  We denote the complex scalar in each half-hypermultiplet as $Q^i_a$ (where the $Sp(N)$ flavour indices take values $i,j = 1, \ldots, 2N$).  

\begin{figure}[ht]
\begin{center}
  \includegraphics[totalheight=4cm]{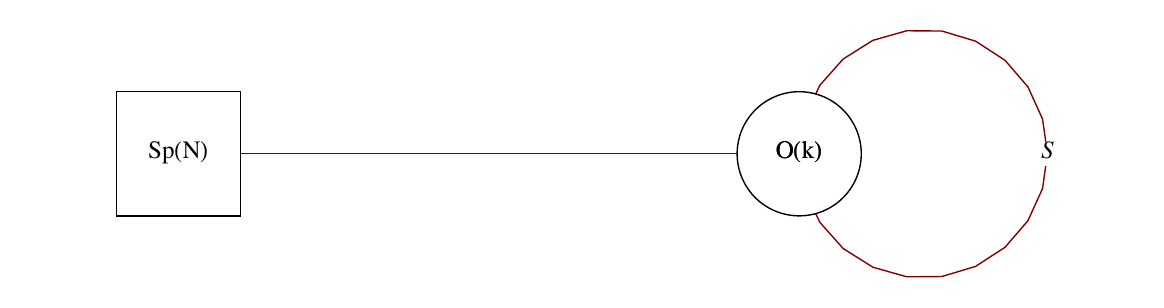}
 \caption{The $\cN=2$ quiver diagram for $k$ $Sp(N)$ instantons on $\BC^2$.  The circular node represents the $O(k)$ gauge symmetry and the square node represents the $Sp(N)$ flavour symmetry. The line connecting the $Sp(N)$ and $O(k)$ groups denotes $2kN$ half-hypermultiplets, and the loop around the $O(k)$ group denotes the second rank (reducible) symmetric tensor.}
  \label{f:N2SpN}
\end{center}
\end{figure}

\begin{figure}[ht]
\begin{center}
  \includegraphics[totalheight=6cm]{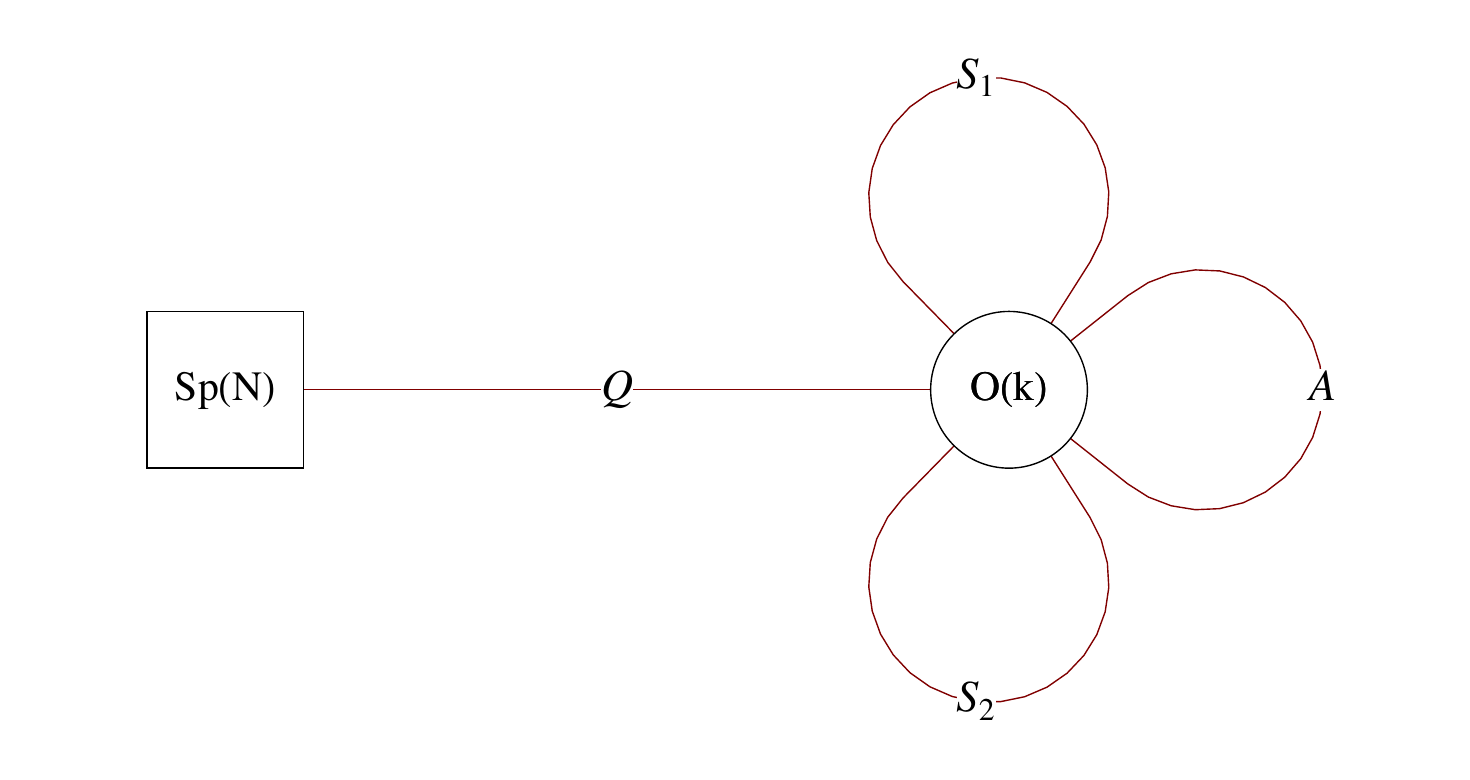}
 \caption{The $\cN=1$ quiver (flower) diagram for $k$ $Sp(N)$ instantons on $\BC^2$, with $A$ being an antisymmetric tensor (adjoint field) and $S_1, S_2$ being symmetric tensors of $Sp(k)$.  The superpotential is $W = Q \cdot A \cdot Q + \epsilon_{\alpha \beta} S_\alpha \cdot A \cdot S_\beta$.}
  \label{f:N1SpN}
\end{center}
\end{figure}

The $\mN=2$ and $\mN=1$ quiver diagrams are  given respectively in \fref{f:N2SpN} and \fref{f:N1SpN}.  The $\cN=1$ superpotential is
\bea
W &=&  Q \cdot A \cdot Q +\left( S_1 \cdot A  \cdot S_2 - S_2 \cdot A \cdot S_1 \right) \nn \\
&=& Q \cdot A \cdot Q + \epsilon_{\alpha \beta} S_\alpha \cdot A \cdot S_\beta~,
\label{WSOk}
\eea
where we have suppressed the contractions over the flavour indices by the tensor $J_{ij}$ (an invariant tensor of $Sp(N)$) and the contractions over the gauge indices by $\delta^{ab}$ (an invariant tensor of $SO(k)$).  The epsilon tensor $\epsilon_{\alpha \beta}$ in the second line is an invariant tensor of the global 
$SU(2)$ symmetry which interchanges $S_1$ and $S_2$.
The mass term transforming in the adjoint of $Sp(N)$ is set to 0.

Let us compute the quaternionic dimension of the Higgs branch.  From the $\cN=2$ quiver diagram, the lines connecting the $Sp(N)$ and $O(k)$ groups denotes $2kN$ half-hypermultiplets (equivalently, $kN$ hypermultiplets), and the loop around the $O(k)$ group gives $\frac{1}{2}k(k+1)$ hypermultiplets.  Hence, we have in total $kN+\frac{1}{2}k(k+1)$ degrees of freedom.    On the Higgs branch, we assume that $O(k)$ is completely broken and hence there are $\frac{1}{2}k(k-1)$ broken generators.  As a result of the Higgs mechanism, the vector multiplet gains $\frac{1}{2}k(k-1)$ degrees of freedom and becomes massive.  Hence, the $\left[ kN+\frac{1}{2}k(k+1) \right] - \frac{1}{2}k(k-1)= k(N+1)$ degrees of freedom are left massless.  Thus, the Higgs branch is $k(N+1)$ quaternionic dimensional or $2k(N+1)$ complex dimensional:
\bea
\dim_{\BC} \MH_{k,N} = 2k(N+1) = 2kh_{Sp(N)}~, \label{SpNdim}
\eea
where $h_{Sp(N)} = N+1$ is the dual coxeter number of the $Sp(N)$ gauge group.

We list the charges and the representations under which the fields transform in \tref{t:chargeSp}.
\begin{table}[htdp]
\begin{center}
\begin{tabular}{|c|c|c|c|c|c|}
\hline
Field & $SO(k)$ &$Sp(N)$& $SU(2)_g$ global & $U(1)$ global\\
\hline
Fugacity: & $z_1, \ldots, z_{k}$& $x_1, \ldots, x_{\lfloor N/2 \rfloor}$ & $x$ & $t$ \\
\hline
$A$ & $[0,1,\ldots,0]$  &  $[0,\ldots,0]$  & [0] &  0\\
$S_1, S_2$ & $[2,0,\ldots,0]+[0, \ldots, 0]$  &  $[0,\ldots,0]$  & [1] &  1 \\
$Q$ & $[1,0, \ldots, 0]$  & $[1,0, \ldots, 0]$  & [0] &  1\\ 
\hline
\end{tabular}
\caption{The charges and the representations under which various fields transform.  The fugacites of each field are assigned according to this table.}
\label{t:chargeSp}
\end{center}
\end{table}%

\subsubsection{One $Sp(N)$ instanton on $\BC^2$: $k=1$}
For $k=1$, the gauge group becomes $O(1) \cong \BZ_2$.  Recall that we have $2N$ hypermultiplets $Q^i$ and two gauge singlets $S_1$ and $S_2$.  It is then easy to see that the moduli space in this case is
\bea
\MH_{k=1,N} = \BC^{2N} /\BZ_2 \times \BC^2~,
\eea
where the factor $\BC^2$ is parametrised by $S_1$ and $S_2$, the $\BC^{2N}$ is parametrised by $Q^i$, and the orbifold action $\BZ_2$ is $-1$ on each coordinate of $\BC^{2N}$.  Observe that $\MH_{k=1,N}$ is $2(N+1)$ complex dimensional, in accordance with \eref{SpNdim}.
Physically, the $\BC^2$ corresponds to the position (4 real coordinates) of the instanton. 
The coherent component of the one $Sp(N)$ instanton moduli space is therefore $\BC^{2N} /\BZ_2$.

One can see the last statement clearly from the Hilbert series.  The Hilbert series of $\BC^{2N} /\BZ_2$ is given by the discrete Molien formula (see, \emph{e.g.}, \cite{pleth1}):
\bea
g(t, x_1, \ldots, x_{N}; \BC^{2N} /\BZ_2) &=& \frac{1}{2} \left( \PE \left[ [1,0,\ldots,0]_{Sp(N)} t \right] + \PE \left[ [1,0,\ldots,0]_{Sp(N)} (-t) \right] \right) \nn \\
&=& \sum_{k=0}^\infty [2k, 0, \ldots, 0] t^{2k}~, \label{gensp}
\eea
where the plethystic exponential can be written explicitly as
\bea
\PE \left[ [1,0,\ldots,0]_{Sp(N)} t \right] = \frac{1}{\prod_{a=1}^N (1-t x_a)(1-t/x_a)} = \sum_{n=0}^\infty  \left[n,0,\ldots,0 \right ]_{Sp(N)} t^n ,
\nn
\eea
and the $\BZ_2$ acts on $t$ by projecting to even powers.
The last equality of \eref{gensp} follows from the fact that the plethystic exponential generates symmetrisation.
This is indeed the Hilbert series for the coherent component of the one $Sp(N)$ instanton moduli space. The choice of $x_a$ in this formula is not the natural choice of weights in the representation but rather a linear combination of weights which is convenient for writing this particular formula. 

\section{$\mN=2$ Supersymmetric $SU(N_c)$ gauge theory with $N_f$ flavours}

\begin{figure}[htbp]
\begin{center}
\includegraphics[totalheight=1.7cm]{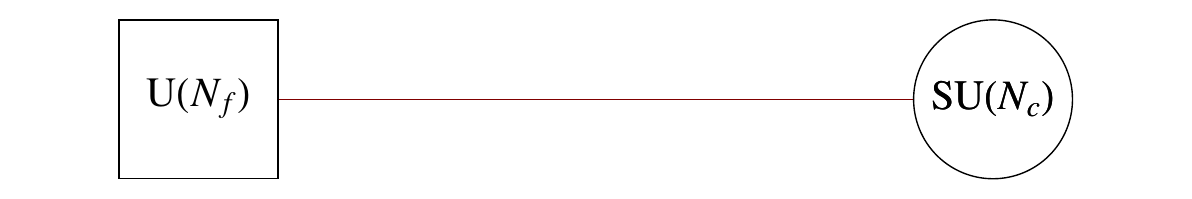}
\caption{$\mN=2$ quiver diagram for $SU(N_c)$ gauge theory with $N_f$ flavours.}
\label{f:N2SUNcwithNf}
\end{center}
\end{figure}

This section deals with the computation of the Hilbert series for the Higgs branch of the $\cN=2$ $SU(N_c)$ supersymmetric gauge theory with $N_f$ flavours. It serves as a preparation for the discussion in Section \sref{EN}, were the results will be used in checking Argyres-Seiberg duality.
The global symmetry of this theory is $U(N_f) = U(1)_B \times SU(N_f)$ and since it plays a crucial role on the Higgs branch this theory will sometimes be called the $U(N_f)$ theory. 
The special case of $N_c =2$ and $N_f = 4$ is discussed in \sref{onesonins} and is revisited below.
The $\cN=2$ quiver diagram for this theory is depicted in \fref{f:N2SUNcwithNf}.

\begin{figure}[htbp]
\begin{center}
\includegraphics[totalheight=3.5cm]{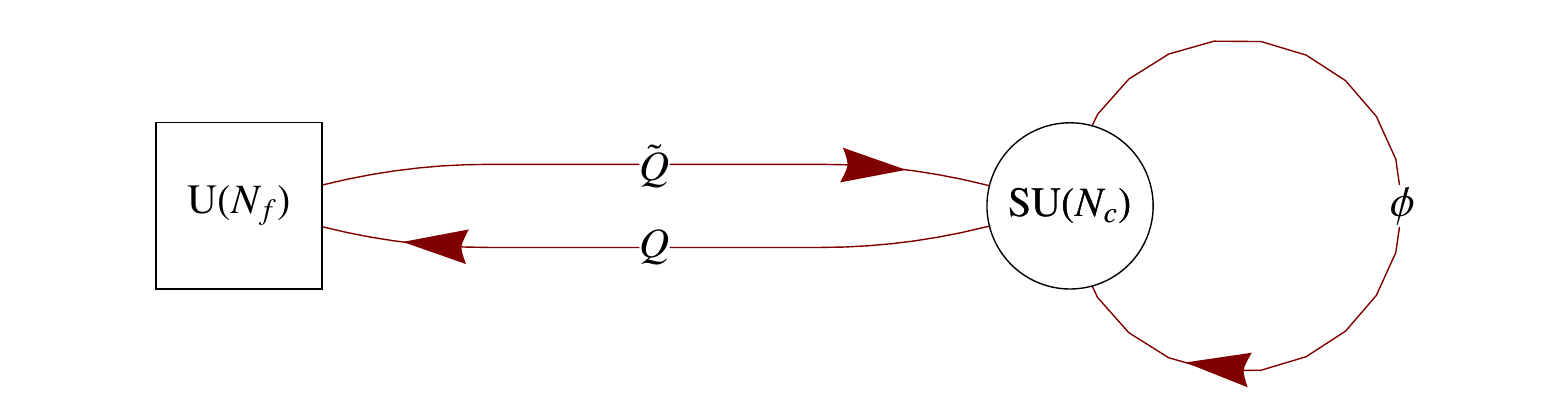}
\caption{$\mN=1$ quiver diagram for $SU(N_c)$ gauge theory with $N_f$ flavours. The superpotential is $W = \tQ \cdot \phi \cdot Q$.}
\label{f:N1SUNcwithNf}
\end{center}
\end{figure}

The $\cN=1$ quiver diagram is depicted in \fref{f:N1SUNcwithNf} and the superpotential after setting the masses to 0 is given by
\beq
W = \tQ \cdot \phi \cdot Q,
\eeq
giving the F-term equations on the Higgs branch, $\phi=0$ and $Q\tQ  = 0$, where the last equation has only $N_c^2-1$ equations and not $N_c^2$. The trace meson $\tQ \cdot Q$ need not vanish.

The Higgs branch of this theory has a Hilbert series which is easy to write down as an integral over the Haar measure of $SU(N_c)$. The reason for this lies partly with supersymmetry and partly with the simplicity of the gauge and matter content. We first argue that the F-flat moduli space is a complete intersection. Since the quaternionic dimension of the Higgs branch is $N_c N_f - (N_c^2-1)$, the complex dimension of the F-flat moduli space is expected to be $N_c^2-1$ higher than this one. Adding these together, we get that the complex dimension of the F-flat moduli space is $2N_c N_f - (N_c^2 - 1)$. On the other hand, these are precisely the number of degrees of freedom. There are $2N_c N_f$ complex variables which are subject to $N_c^2-1$ equations on the Higgs branch. We therefore conclude that the F-flat moduli space is a complete intersection and its Hilbert series can be written as a ratio of two plethystic exponentials,
\beq \label{ffncnf}
\gf_{N_c, N_f} = \frac{\PE \left [ [1,0,\ldots, 0]_{SU(N_c)} [0,\ldots, 0,1]_{SU(N_f)} t_1 + [0,\ldots, 0,1]_{SU(N_c)} [1,0,\ldots, 0]_{SU(N_f)}{t_2}\right ]}{\PE \left [ [1,0, \ldots, 0,1]_{SU(N_c)} t^2\right ]}~,  
\eeq
where $t_1=t b$ and $t_2={t}/b$ are respectively the global $U(1)$ fugacities for $Q$ and $\tQ$ and $b$ is the fugacity for the baryonic symmetry $U(1)_B$.
The Higgs branch is given by integrating this Hilbert series using the $SU(N_c)$ Haar measue,
\beq
\gH_{N_c,N_f} = \int d\mu_{SU(N_c)}  \gf_{N_c, N_f}~. \label{gHNcNf}
\eeq



\subsection{The case of $N_c=3$ and $N_f =6$}
In this subsection, we focus on the $\mN=2$ supersymmetric $SU(3)$ gauge theory with $6$ flavours.  

From \eref{ffncnf}, the F-flat Hilbert series after setting all $U(6)$ fugacities to 1 can be written as
\bea
\gf_{N_c=3, N_f=6} = \frac{\left(1-t^2\right)^2 \left(1-\frac{t^2 z_1}{z_2^2}\right) \left(1-\frac{t^2}{z_1 z_2}\right) \left(1-\frac{t^2 z_1^2}{z_2}\right) \left(1-\frac{t^2 z_2}{z_1^2}\right) \left(1-t^2 z_1 z_2\right) \left(1-\frac{t^2 z_2^2}{z_1}\right)}{\left(1-t z_1\right)^6 \left(1-t z_2\right)^6 \left(1-\frac{t}{z_1}\right)^6 \left(1-\frac{t}{z_2}\right)^6 \left(1-\frac{t z_1}{z_2}\right)^6 \left(1-\frac{t z_2}{z_1}\right)^6}~,\nn \\
\eea
where $z_1$ and $z_2$ are the $SU(3)$ fugacities. 
The Haar measure for $SU(3)$ is
\bea
\int  \ud \mu_{SU(3)} &=& \frac{1}{(2\pi i)^2} \oint_{|z_1| =1} \frac{dz_{1}}{z_{1}} \oint_{|z_2| =1} \frac{dz_{2}}{z_{2}}  \left(1-z_{1}z_{2}\right)\left(1-\frac{z^{2}_{1}}{z_{2}}\right)\left(1-\frac{z^{2}_{2}}{z_{1}}\right)~,
\eea
After integrating over $z_1$ and $z_2$, we obtain the Hilbert series:\footnote{In using the residue theorem, the non-trivial contributions to the first integral over $z_1$ come from the poles $z_1 = t, ~ t z_2$, and the non-trivial contributions to the second integral over $z_2$ come from the poles $z_2 = t, ~ t^2$.}
\begin{equation}\label{HSU6}
\gH_{N_c =3, N_f =6} (t) = \frac{P(t)}
{(1-t)^{20} (1+t)^{16} (1+t+t^2)^{10}}~,
\end{equation}
where the numerator $P(t)$ is a palindromic polynomial of degree $36$:
\bea
P(t) &=& 1+6 t+41 t^2+206 t^3+900 t^4+3326 t^5+10846 t^6+31100 t^7+79677 t^8+\nn \\ 
&& + 183232 t^9+381347 t^{10}+720592 t^{11} +1242416 t^{12}+1959850 t^{13}+ \nn \\
&& + 2837034 t^{14}+3774494 t^{15} +4624009 t^{16}+5220406 t^{17}+5435982 t^{18} \nn \\
&& +\ldots ~\text{(palindrome)}~ \ldots + t^{36}~.
\eea
Note that the space is $20 = 2(3\cdot6-8)$ complex-dimensional, as expected.  
The first few orders of the power expansion of \eref{HSU6} reads
\bea
\gH_{N_c =3, N_f =6} (t)  = 1+36 t^2+40 t^3+630 t^4+1120 t^5+\ldots~. \label{unrefine36}
\eea
The plethystic logarithm is
\bea
PL \left[ \gH_{N_c =3, N_f =6} (t) \right] = 36 t^2 + 40 t^3 - 36 t^4 - 320 t^5 - 435 t^6 + \ldots~. 
\eea

\paragraph{The fully refined Hilbert series.}
In fact, one can obtain the fully refined Hilbert series directly from \eref{ffncnf} and \eref{gHNcNf}.  The result can be written as a power series 
\bea
&& \gH_{N_c =3, N_f =6}(t_1, t_2; x_1, x_2, x_3, x_4, x_5) =  \nn \\
&& \frac{1}{1-t_1 t_2} \sum_{n_1=0}^\infty \sum_{n_2=0}^\infty \sum_{n_3=0}^\infty \sum_{n_4=0}^\infty [n_1, n_2, n_3 + n_4, n_2, n_1]_{SU(6)} {t_1}^{ n_1 + 2 n_2 + 3 n_3 }  {t_2}^{ n_1 + 2 n_2 + 3 n_4}~. \qquad \label{refu6sum}
\eea
where $x_1, \ldots, x_5$ are the $SU(6)$ fugacities. 

The plethystic logarithm of \eref{refu6sum} is
\bea
&& \PL \left[ \gH_{N_c =3, N_f =6} (t_1, t_2; x_1, x_2, x_3, x_4, x_5) \right] =( [0,0,0,0,0]+[1,0,0,0,1]) t_1 t_2 + \nn \\
&& \quad [0,0,1,0,0] (t_1^3 + t_2^3 ) - ( [0,0,0,0,0]+[1,0,0,0,1]) t_1^2 t_2^2 + \ldots~,
\eea
where the gauge invariant operators in the representation $[0,0,0,0,0]+[1,0,0,0,1]$ of $SU(6)$ can be identified as {\it mesons} (see \eref{mesons}) and the operators in the representation $[0,0,1,0,0]$ of $SU(6)$ can be identified as {\it baryons} and {\it antibaryons} (see \eref{baryons}).


\subsection{Generalisation to the case $N_f = 2N_c$}
The formula \eref{refu6} can be generalised to the case $N_f = 2N_c$.  Let us first consider the simplest case of: $N_f = 2N_c = 4$, discussed in \sref{onesonins}.

\paragraph{The $N_c=2$ and $N_f = 4$ case.}  
From \eref{gHso8}, the Hilbert series of the coherent component of the Higgs branch is 
\bea
\gH_{N_c=2, N_f =4} (t; x_1, x_2, x_3, x_4)= \sum_{k=0}^\infty [0,k,0,0]_{SO(8)} t^{2k}~,
\label{su24flv}
\eea
The branching rule of the representation $[0,k,0,0]$ of $SO(8)$ to the subgroup $SU(4) \times U(1)_B$ is given by
\bea \label{so8branch}
[0,k,0,0]_{SO(8)} = \sum_{n_1=0}^\infty \sum_{n_2 = 0}^\infty \sum_{n_3 = 0}^\infty \sum_{n_4 = 0}^\infty [n_1, n_2+n_3, n_1]_{SU(4)} b^{2n_2-2n_3} \delta(k - n_1 - n_2 - n_3 - n_4)~, \qquad
\eea
or equivalently the decomposition identity
\bea \label{so8branch1}
\sum_{k=0}^\infty [0,k,0,0]_{SO(8)} t^{2k} = \frac{1}{1-t^2}\sum_{n_1=0}^\infty \sum_{n_2 = 0}^\infty \sum_{n_3 = 0}^\infty [n_1, n_2+n_3, n_1]_{SU(4)} b^{2n_2-2n_3} t^{2n_1+2n_2+2n_3}~, \qquad
\eea
where $b$ is the fugacity of $U(1)_B$.  Substituting \eref{so8branch} into \eref{su24flv}, we obtain
\bea \label{gH24}
\gH_{N_c=2, N_f =4} (t; x_1, x_2, x_3; b) &=& \sum_{n_1=0}^\infty \sum_{n_2 = 0}^\infty \sum_{n_3 = 0}^\infty [n_1, n_2+n_3, n_1] t^{2n_1+2n_2+2n_3+2n_4} b^{2n_2-2n_3} \nn \\
&=& \frac{1}{1-t_1 t_2} \sum_{n_1=0}^\infty \sum_{n_2 = 0}^\infty \sum_{n_3 = 0}^\infty [n_1, n_2+n_3, n_1] t_1^{n_1+2n_2} t_2^{n_1+2n_3}, \qquad
\eea 
where in the last line we take $t_1 = t b$ and $ t_2 = t b^{-1}$.

\paragraph{Generalisation.}  From \eref{refu6sum} and \eref{gH24}, we conjecture that the Hilbert series for the Higgs branch of the $SU(N_c)$ gauge theory with $N_f = 2N_c$ flavours can be written in terms of $SU(2N_c)$ representations as
\bea
&& \gH_{N_f =2N_c}(t_1, t_2; x_1, \ldots, x_{2N_c-1}) =  \nn \\
&& \frac{1}{1-t_1 t_2} \sum_{n_1 = 0}^\infty  \cdots \sum_{n_{N_c+1} =0}^\infty [n_1, n_2, \ldots, n_{N_c-1} , n_{N_c} + n_{N_c+1}, n_{N_c-1}, \ldots, n_2, n_1] t_1^{d + N_c n_{N_c}}  t_2^{d + N_c n_{N_c+1}}~, \nn \\ \label{refu6}
\eea
where $d={ \sum_{k=1}^{N_c-1} k n_k}$.  This formula can be checked by plugging in the dimensions of the representations, one finds that the Higgs branch is $2(N_c^2+1)$ complex dimensional, as expected.  Note the similarity between \eref{refu6} and the Hilbert series of $\cN=1$ SQCD (see (5.1) of \cite{apercu}); however, they are not identical  -- the moduli space of $\cN=1$ SQCD with $N_f \geq N_c$ is $2N_cN_f - (N_c^2-1)$ complex dimensional, whereas the moduli space of the $\cN=2$ gauge theory is $2N_c N_f - 2(N_c^2-1)$ complex dimensional.  

The plethystic logarithm of \eref{refu6} indicates that:
\bi 
\item At the order $t_1 t_2$, there are gauge invariants transforming in the representation $[0,\ldots,0]+[1,0,\ldots,0,1]$ of $SU(N_f)$ and carrying $U(1)_B$ charge $0$  These operators are {\it mesons}: 
\bea
M^i_j = Q^i_a \tQ^a_j~, \label{mesons}
\eea
where $a=1, \ldots, N_c$ and $i, j =1, \ldots, N_f$. 
\item At the order $t_1^{N_c}$ and $t_2^{N_c}$, there are gauge invariants transforming in the representation $[0,\ldots, 0,1, 0, \ldots 0]$ of $SU(N_f)$ and carrying $U(1)_B$ charges $N_c$ and $-N_c$.  These operators are respectively {\it baryons} and {\it antibaryons}:
\bea
B^{i_1, \ldots, i_{N_c}} &=& \epsilon^{a_1 \ldots a_{N_c}} Q^{i_1}_{a_1} \ldots Q^{i_{N_c}}_{a_{N_c}}~, \nn \\
\tB_{i_1, \ldots, i_{N_c}} &=& \epsilon_{a_1 \ldots a_{N_c}} \tQ_{i_1}^{a_1} \ldots \tQ_{i_{N_c}}^{a_{N_c}}~.  \label{baryons}
\eea
\ei
These generators are indeed identical to those of the $\cN=1$ SQCD.  Hence, they satisfy the relations given by (3.11) and (3.12) of \cite{apercu}:
\bea \label{cons1}
(*B)\widetilde{B} &=& *(M^{N_c}) ~, \nn \\
M \cdot *B &=& M \cdot *\widetilde{B} =0 ~.
\eea
where $(*B)_{i_{N_c+1} \ldots i_{N_f}} = \frac{1} {N_c !} \epsilon_{i_1 \ldots i_{N_f}} B^{i_1 \ldots i_{N_c}}$ and a `$\cdot$' denotes a contraction of an upper with a lower flavour index.
In addition, the F-terms impose further relations.  These are given by (2.23) and (2.24) of \cite{Argyres:1996eh}:
\bea
M' \cdot B = \tB \cdot M' &=& 0~, \nn \\
M \cdot M' &=& 0~,
\eea
where 
\bea
(M')^i_j = M^i_j - \frac{1}{N_c} (\tr M) \delta^i_j~.
\eea

\section{Exceptional groups and Argyres-Seiberg dualities}\label{EN}
In this section, we consider the Hilbert series of a single $G$ instanton on $\BR^4$ where $G$ is one of the 5 exceptional groups. 
It is shown that the conjecture is consistent with the dimension of the instanton moduli space, by explicitly summing the unrefined Hilbert series. In the cases of $E_6$ and $E_7$, we also check that the proposed Hilbert Series are consistent with Argyres-Seiberg dualities found in  \cite{Argyres:2007cn, Argyres:2007tq, Gaiotto:2008nz, Gaiotto:2009we, Gaiotto:2009gz, Tachikawa:2009rb,  Benini:2009gi}. Only for the case of $E_6$, we are able to carry out a full all-order check. In the case of $E_7$, we just match the lower dimension BPS operators.   Notice that the check for BPS operators of scaling dimension $2$ is equivalent to the check that the symmetries on both sides of the duality are the same. This is because  BPS operators of scaling dimension $2$ are in the same super multiplet of the flavour currents.

\paragraph{Notation:} In this section, when there is no ambiguity, we denote special unitary ($SU$) groups in the quiver diagrams by their ranks.   Each $U(1)$ global symmetry is associated with a hypermultiplet and hence each solid line connecting two nodes represents a $U(1)$ global symmetry.  The dashed lines are not associated with bi-fundamental hypermultiplets and do not correspond to $U(1)$ global symmetries. Square nodes with an index 1 do not count as a $U(1)$ global symmetry.

\subsection{$E_6$}
The Hilbert series of one $E_6$-instanton on $\BR^4$ is given by \eref{HS}:
\beq\label{E6count0}
\HS_{E_6} (t; x_1,\ldots, x_6) = \sum_{k=0}^\infty [0, k, 0, 0, 0, 0] t^{2k} .
\eeq
By setting the $E_6$ fugacities to 1, this equation can be resumed and written in the form of \eref{generalHS}:
\beq\label{E6count}
\HS_{E_6} (t; 1, \ldots, 1)=  \frac{P_{E_6} (t)}{(1-t^2)^{22}},
\eeq
where 
\bea
P_{E_6} (t) &=& (1+t^2)(1+55 t^2+ 890 t^4 + 5886 t^6 + 17929 t^8 + 26060 t^{10} + \nn \\
&&  \ldots~\text{(palindrome)}~ \ldots + t^{20})~.
\eea
This confirms that the complex dimension of the moduli space is $2h_{E_6}-2=22$, where $h_{E_6}=12$ is the dual Coxeter number of $E_6$.

\subsubsection{Duality between the $6-\bullet-2-1$ quiver theory and the $SU(3)$ gauge theory with 6 flavours}
\begin{figure}[htbp]
\begin{center}
\includegraphics[totalheight=5cm]{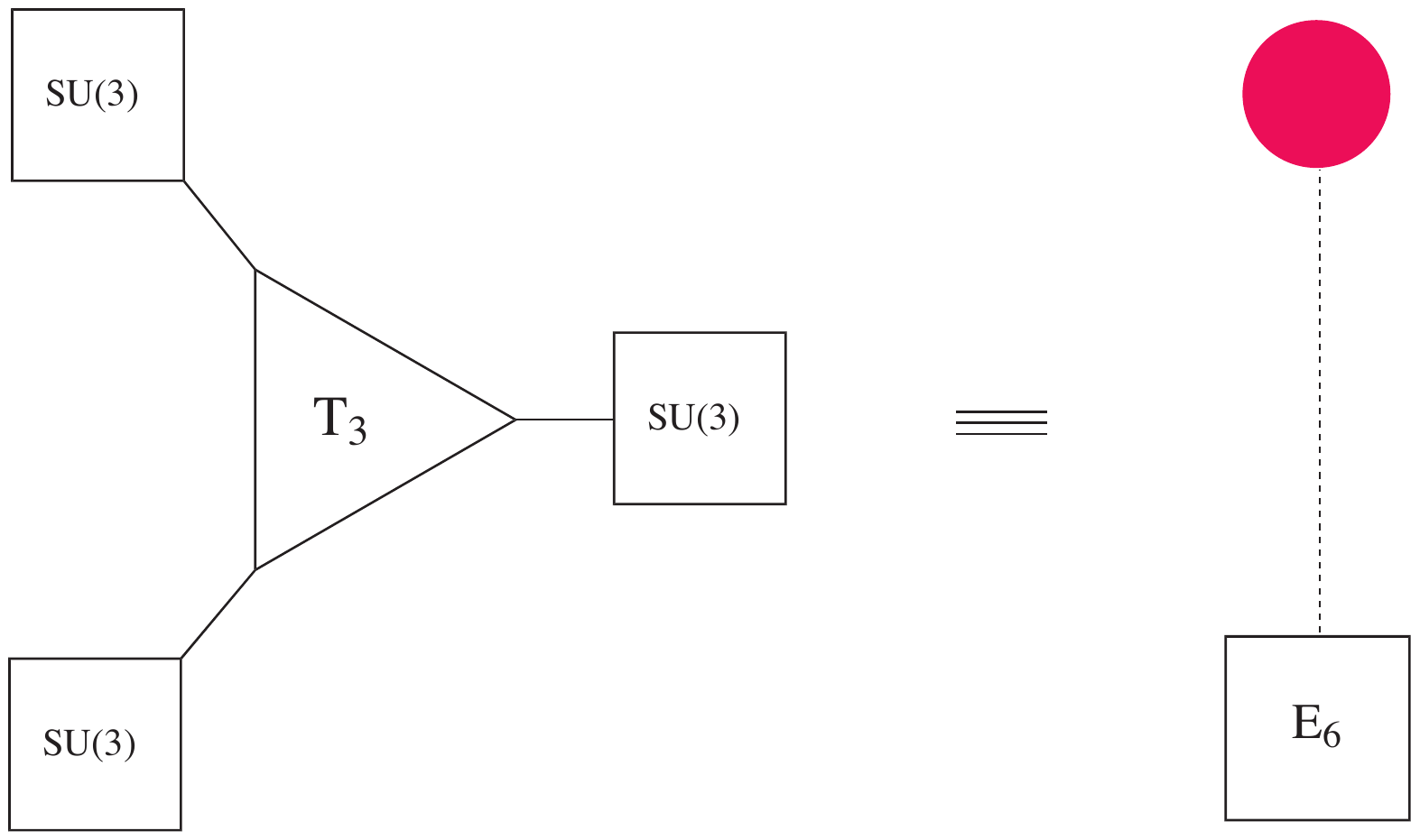}
\caption{{\bf Left:} The $E_6$ theory arising from $3$ M5-branes wrapping a sphere with $3$ maximal punctures, each is associated to $SU(3)$ global symmetry.  The $SU(3)^3$ symmetry enhances to $E_6$.  {\bf Right:}  The quiver diagram representing the $E_6$ theory.  The red blob denotes a theory with an unknown Lagrangian description.  The $E_6$ global symmetry is indicated in the square node.}
\label{f:T3}
\end{center}
\end{figure}

As discussed in \cite{Gaiotto:2009we},  the strongly interacting SCFT with $E_6$ flavour symmetry can be realised as $3$ M5-branes wrapping a sphere with $3$ punctures. These punctures are of the maximal type, each one is associated to $SU(3)$ global symmetry.  The global symmetry $SU(3)^3$ enhances to $E_6$.  This theory is also known as the $T_3$ theory \cite{Minahan:1996fg, Minahan:1996cj, Gaiotto:2009we, Gaiotto:2009gz} and is denoted by the left picture of \fref{f:T3}.  There is no known Lagrangian description for this theory.  

The $E_6$ theory is denoted by a `quiver diagram' which is analogous to those in previous sections.  This is given in the right picture of \fref{f:T3}.  The red blob denotes a theory with an unknown Lagrangian.  The $E_6$ global symmetry is indicated in the square node. Below it is demonstrated that even though the Lagrangian is not known, it is still possible to make statements about the spectrum of operators for this theory.

\begin{figure}[htbp]
\begin{center}
\includegraphics[totalheight=1.5cm]{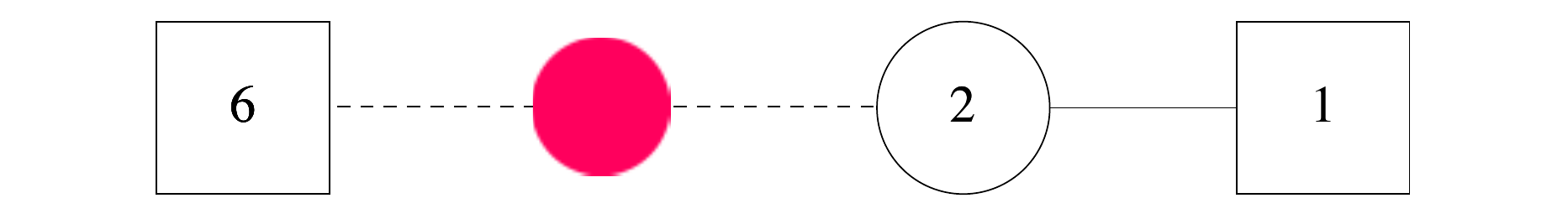}
\caption{ {\bf The $6-\bullet-2-1$ quiver theory:} From the $E_6$ theory, the global symmetry $E_6$ is decomposed into the subgroup $SU(2) \times SU(6)$.   The $SU(2)$ symmetry is gauged and is coupled to the $2-1$ tail. The $U(1)$ global symmetry is associated with the solid line in the quiver diagram. The flavour symmetry is $SU(6) \times U(1)$.}
\label{f:E6af}
\end{center}
\end{figure}

\begin{figure}[htbp]
\begin{center}
\includegraphics[totalheight=1.5cm]{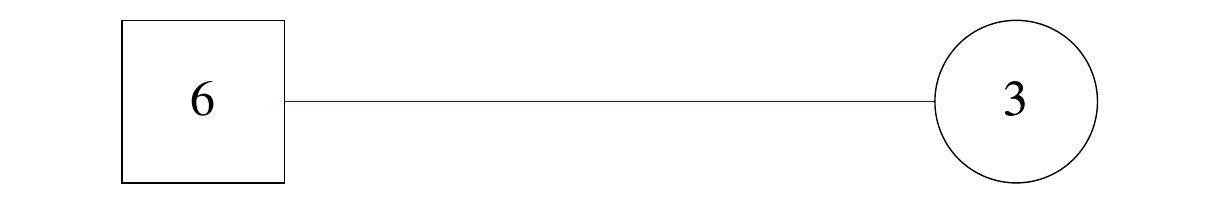}
\caption{The $SU(3)$ gauge theory with 6 flavours.  This theory is conjectured by Argyres-Seiberg to be dual to the $6-\bullet-2-1$ quiver theory.}
\label{f:E6be}
\end{center}
\end{figure}

The $E_6$ theory can be used to construct a quiver gauge theory called the $6-\bullet-2-1$ theory, depicted in \fref{f:E7af}.  This theory is proposed by Argyres and Seiberg \cite{Argyres:2007cn} to be dual to an $SU(3)$ gauge theory with 6 flavours, whose quiver diagram is shown in \fref{f:E6be}.  The appearance of the tail in \fref{f:E6af} seems to be a generic feature of these dualities and follows from the splitting of branes when ending on the same brane - see Figure 20 of \cite{Hanany:1996ie}. 

Let us summarise a construction of the $6-\bullet-2-1$ quiver theory.  The global symmetry $E_6$ can be decomposed into the subgroup $SU(2) \times SU(6)$. The $SU(2)$ symmetry is gauged and is coupled to the $2-1$ tail, as depicted in \fref{f:E6af}. The resulting theory is the {\bf the $6-\bullet-2-1$ quiver theory}.  The $U(1)$ global symmetry is associated with the solid line in the quiver diagram. The global symmetry is thus $SU(6) \times U(1) \cong U(6)$.

Note that a necessary condition for two theories to be dual is that they have the same global symmetry. Indeed, both of the $6-\bullet-2-1$ quiver theory and the $SU(3)$ gauge theory with 6 flavours have the same global symmetry $U(6)$, even though these symmetries arise from different sources in each case.





\paragraph{A branching rule for $E_6$ to $SU(2) \times SU(6)$.}  To proceed, we first decompose the $E_6$ representations into representations of $SU(2)\times SU(6)$. For this it is useful to introduce the fugacity map. The fugacities $u_1, u_2, \ldots, u_6$ of $E_6$ can be mapped to the fugacities $x$ of $SU(2)$ and $y_1, \ldots, y_5$ of $SU(6)$ as follows:
\bea
u_1 = x y_5,\quad u_2 = y_1 y_5,\quad u_3 = y_5^2,\quad u_4 = y_2 y_5^2,\quad u_5 = y_3 y_5,\quad u_6 = y_4~.
\eea
Using this map, one can decompose the character of an $E_6$ representation into the characters of $SU(2) \times SU(6)$ representations.  For example, if we denote a representation of $SU(2)\times SU(6)$ of highest weight $m$ for $SU(2)$ and highest weights $n_1, \ldots, n_5$ for $SU(6)$ by $[m; n_1, \ldots, n_5]$, then one finds that
\bea
\left[0,1,0,0,0,0 \right]_{E_6} & = & [0; 1, 0, 0, 0, 1]+[1; 0, 0, 1, 0, 0]+[2; 0, 0, 0, 0, 0] , \nn \\
\left[0,2,0,0,0,0 \right]_{E_6}& = & [0; 0, 0, 0, 0, 0]+[0; 0, 1, 0, 1, 0]+[0; 2, 0, 0, 0, 2] +   \nn \\ 
&& [1; 0, 0, 1, 0, 0] +  [1; 1, 0, 1, 0, 1] + [2; 1, 0, 0, 0, 1]+ \nn \\
&& [2; 0, 0, 2, 0, 0]+[3; 0, 0, 1, 0, 0]+[4; 0, 0, 0, 0, 0]~ . \label{low26decomd}
\eea
These equalities can be checked by matching the characters of the representations on both sides.  
The general formula for the decompositions of $Adj^k$ for any $k$ is given in \eref{branchinggen}.

The decompositions \eref{low26decomd} can be written in terms of dimensions as
\bea
78 & \rightarrow& (1,35) \oplus (2,20) \oplus (3,1) \nn \\
2430 & \rightarrow & (1,1) \oplus (1,189) \oplus (1,405) \oplus (2,20) \oplus (2,540) \oplus (3,35) \oplus \nn \\
&&  (3,175) \oplus (4,20) \oplus (5,1)~.
\eea

\paragraph{Counting BPS operators of the $SU(3)$ gauge theory with  6 flavours.}
In what follows, starting from (\ref{E6count0}), we count BPS operators in the $SU(3)$ gauge theory with $6$ flavours by computing the $SU(2)$ gauge invariant spectrum.  For now, let us first do this order by order for the operators of small scaling dimensions.  In the later subsections, we present a method to count the operators to all orders.

\bi
\item At level $t^2$, we expect the $35$ to survive, as it is an $SU(2)$ singlet. Denote the $2-1$ hypermultiplet in \fref{f:E6af} by $q$ and $\qt$. Set $q$ to have fugacity $t b^3$ and $\qt$ to have fugacity $t/b^3$, where the normalization 3 is chosen for matching with the $U(6)$ baryons. One can construct another $SU(2)$ invariant which is a singlet under $SU(6)$, by forming $q \qt$. We therefore expect the SU(3) theory with 6 flavours to have $35_0+1_0$ at order $t^2$, where the subscript $0$ refers to the $U(1)_B$ baryonic charge. Indeed, in the $SU(3)$ theory of \fref{f:E6be} these are formed by the $SU(3)$ mesons $\tilde Q Q$ that decompose as $35_0+1_0$.

\item At level $t^3$, the $(2,20)$ coupled to $q$ or to $\qt$, leads to the $SU(2)$ invariant operators which transform as $20_3\oplus20_{-3}$. This contributes the term $20 (b^3+1/b^3)t^3$ to the $U(6)$ Hilbert series.

\item At level $t^4$ we have the singlets $1+189+405$, and the $35$ from order $t^2$ multiplied by the $SU(6)$-singlet $q \qt$, for a total of $630$ operators.
\ei
These are precisely the first few terms of the Hilbert series \eref{unrefine36} of the Higgs Branch of $SU(3)$ theory with $6$ flavours:
\begin{equation}
\gH_{N_c =3, N_f =6} (t)  = 1 + 36 t^2 + 20(b^3+b^{-3}) t^3 + 630 t^4 + \ldots~.
\end{equation}

\subsubsection{Branching formula for $Adj^k$ of $E_6$ to $SU(2) \times SU(6)$}
In this subsection, we carry out the decomposition of the $Adj^k$-irreducible representations of $E_6$ into $SU(2) \times SU(6)$ to all order in $k$. This gives a useful check of the Argyres-Seiberg duality to all orders. The general form of the decomposition is as follows:
\beq
[0, k, 0, 0, 0, 0]_{E_6} = \sum_{m=0}^{2k} [m]_{SU(2)} C^k_m
\eeq
where $C^k_m$ is a reducible representation of $SU(6)$. 
The sets of irreps of $SU(6)$ entering in $C^k_m$ is constructed starting by the representation $R^L_p$, defined by:
\begin{eqnarray}\nonumber
R^L_{p > 0} & = &\sum_{n=0}^L \sum_{i+2j+3/2k=n} \left[ i,j,k+p,j,i\right]\\
R^{2L}_{p = 0} &=& \sum_{n=0}^L \sum_{i+2j+3/2k=2n} \left[ i,j,k,j,i\right]\\
R^{2L+1}_{p = 0} &=& \sum_{n=0}^L \sum_{i+2j+3/2k=2n+1} \left[ i,j,k,j,i\right]\nonumber
\end{eqnarray}
Notice that only $SU(6)$-irreps whose Dynkin labels are symmetric enter the sum, and that $R^k_n$ contains an irreducible representation at most one time. The $C^k$ are given in terms of the $R^L_p$ by
\begin{eqnarray}\nonumber
C^k_{2m} &=& \sum_{j=0}^m R^{k-m-j}_j\\ 
C^k_{2m+1} &=& \sum_{j=0}^m R^{k-m-1-j}_j
\end{eqnarray}
In $C^k_m$ the same irrep can appear multiple times.
Summing these together we find the decomposition identity
\bea \label{branchinggen}
&&\left (1-t^4 \right ) \sum_{k=0}^\infty [0,k,0,0,0,0]_{E_6} t^{2k}  \\ &=& \sum_{n_1=0}^\infty \sum_{n_2=0}^\infty \sum_{n_3=0}^\infty \sum_{n_4=0}^\infty \sum_{n_5=0}^\infty [n_1 + 2 n_2]_{SU(2)}[ n_3, n_4, n_1 + 2 n_5, n_4, n_3]_{SU(6)} t^{2n_1 + 2n_2 + 2n_3 + 4 n_4 + 6 n_5} \nn \\ \nn
&+& \sum_{n_1=0}^\infty \sum_{n_2=0}^\infty \sum_{n_3=0}^\infty \sum_{n_4=0}^\infty \sum_{n_5=0}^\infty [n_1 + 2 n_2+1]_{SU(2)}[ n_3, n_4, n_1 + 2 n_5+1, n_4, n_3]_{SU(6)} t^{2n_1 + 2n_2 + 2n_3 + 4 n_4 + 6 n_5+4} .\nn 
\eea
Using these all order results, we can proceed to refine $\HS_{E_6}(t)$ in (\ref{E6count}) to a function of $z$ and $t$ (denoted as $\HS_{E_6}(z,t)$), where $z$ is the $SU(2)$ fugacity.



\subsubsection{The Hilbert series of the $6-\bullet-2-1$ quiver theory}
As discussed earlier, the $6-\bullet-2-1$ quiver theory can be obtained by first decomposing the $E_6$ into $SU(2) \times SU(6)$, the $SU(2)$ group is then gauged and is coupled as in the $2-1$ quiver.  This process can also be described as a `sewing' of two Riemann surfaces - one with 3 maximal punctures (corresponding to $E_6$) and the other with two simple puctures (corresponding to $U(2) \times U(1)$).  The Hilbert series can be computed in analogy to the AGT relation \cite{Alday:2009aq, Gadde:2009kb} as follows:
\beq\label{integ}
g_{6-\bullet-2-1} (t) =  \int \ud \mu_{SU(2)}(z)~\HS_{E_6}(t,z)~g_{\mathrm{glue}} (t,z)~ g_{2-1} (t,b,z)~,
\eeq
where the Haar measure for $SU(2)$ is given by
\bea
\int \ud \mu_{SU(2)} =  \frac{1}{2 \pi i} \oint dz \frac{1-z^2}{z}~,
\eea
the Hilbert series for the bi-fundmentals connecting the $SU(2)$ and $U(1)$ nodes is
\bea
g_{2-1} (t,b,z) &=& \PE \left [ [1]_{SU(2)}\left (b^3+b^{-3}\right ) t \right ]  \nn\\
&=& \frac{1}{(1-t z b^3)(1-t \frac{z}{b^3})\left(1 -\frac{t b^3}{z} \right)\left(1 -\frac{t}{z b^3} \right)}~,
\eea
and the `gluing factor' which keeps track of the $3$ F-term relations that comes from differentiating the superpotential by the adjoint chiral field of $SU(2)$ is
\bea
g_{\mathrm{glue}} (t,z) &=& \frac{1}{\PE \left [ [2]_{SU(2)} t^2 \right ]} = \left (1-t^2 z^2 \right ) \left (1-t^2\right ) \left(1- \frac{t^2}{z^2} \right)~.
\eea

The product of $g_{\mathrm{fund}} (z,t)$ and $g_{\mathrm{glue}} (z,t)$ can be written for $b=1$ as
\bea
g_{\mathrm{glue}} (t,z) g_{2-1} (t,1,z) &=& \frac{\left (1-t^2 z^2 \right ) (1-t^2) \left(1- \frac{t^2}{z^2} \right)}{(1-t z)^2\left(1 -\frac{t}{z} \right)^2} \nn \\
 &=& \sum_{n=0}^\infty [n] t^n + \sum_{n=0}^\infty [n+1] t^{n+1} + t^2 - 2 \sum_{n=0}^\infty [n] t^{n+4}~.
\eea
If we restore the $b$ dependence, this sum takes the form
\bea
& & g_{\mathrm{glue}} (t,z) g_{2-1} (t,b,z) = 
\nn \\
 &=& \sum_{n=0}^\infty [n] (t b^3)^n+ \sum_{n=0}^\infty [n+1] \left( \frac{t}{b^3}\right)^{n+1}+ t^2 -\sum_{n=0}^\infty [n] t^{n+4}(b^{3n+6}+b^{-3n-6})~.
\eea

From (\ref{integ}), one sees that the integral is computed by summing over two residues, one at $z=t$ and one at $z=t^2$. For $z=t$, the residue is a rational function with denominator $(1-t)^{21}(1+t+t^2)^{21}$.  For $z=t^2$, the residue is a rational function with denominator $ (1-t)^{21}(1+t)^{16}(1+t^2)^{37}(1+t+t^2)^{21}$. Summing these two residues gives precisely the unrefined Hilbert series $\gH_{N_c =3, N_f =6} (t)$ of (\ref{HSU6}).


For the refined Hilbert series, it is better to exchange the integral in (\ref{integ})  with the sums and use the orthonormality relation
\beq
\oint_{|z|=1} \frac{dz (1-z^2)}{2\pi i z} [n] [m] = \delta_{n,m}
\eeq
to confirm that the fully refined Hilbert series coincides with \eref{refu6sum}.

\subsection{$E_7$}
The Hilbert series of one $E_7$-instanton on $\BR^4$ is given by \eref{HS}:
\beq\label{E7count0}
\HS_{E_7} (t; x_1,\ldots, x_6,x_7) = \sum_{k=0}^\infty [k, 0, 0, 0, 0, 0, 0] t^{2k} .
\eeq
By setting the $E_7$ fugacities to 1, this equation can be resumed and written in the form of \eref{generalHS}:
\bea
\HS_{E_7} (t; 1, \ldots, 1)=  \frac{P_{E_7} (t)}{(1-t^2)^{34}}~,
\eea
where the numerator is a palindromic polynomial of degree $17$ in $t^2$,
\bea
P_{E_7} (t) &=& 1+99 t^2+3410 t^4+56617 t^6+521917 t^8+2889898 t^{10}+10086066 t^{12}+ \nn \\
&& 22867856 t^{14}+34289476 t^{16}+ \ldots~\text{(palindrome)}~\ldots+ t^{34}~.
\eea
This is consistent with the fact that the Higgs branch is $2h_{E_7}-2=34$ complex dimensional, where $h_{E_7}=18$ is the dual Coxeter number of $E_7$.


\subsubsection{Duality between the $6-\bullet-3-2-1$ quiver theory and the $2-4-6$ quiver theory} \label{sec:duale7}

\begin{figure}[htbp]
\begin{center}
\includegraphics[totalheight=5cm]{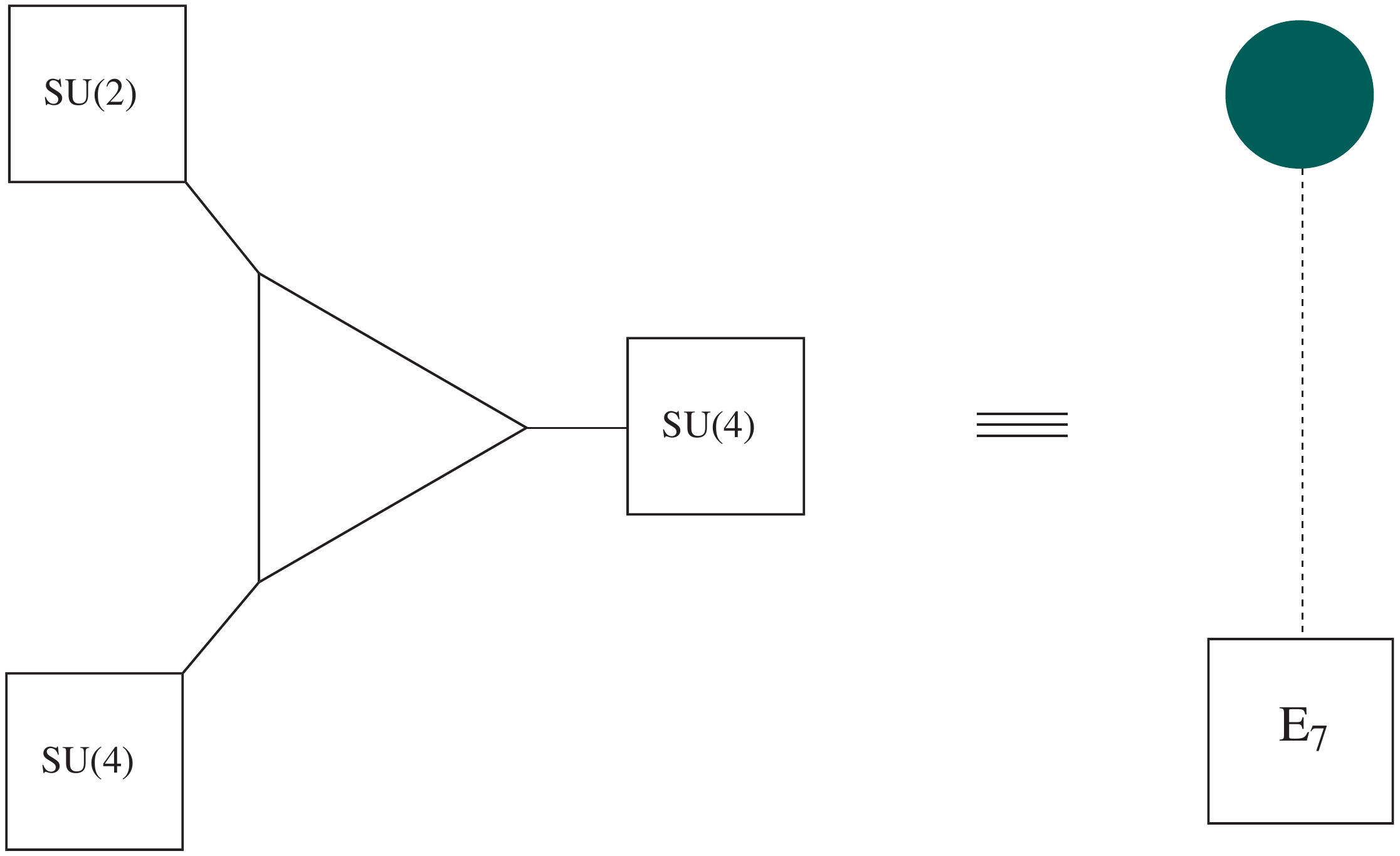}
\caption{{\bf Left:} The $E_7$ theory arising from $4$ M5-branes wrapped over a sphere with $3$ punctures of the type $SU(4)$, $SU(4)$, $SU(2)$. {\bf Right:}  The quiver diagram representing the $E_7$ theory.  The green blob denotes a theory with an unknown Lagrangian description.  The $E_7$ global symmetry is indicated by the square node.}
\label{f:TA3}
\end{center}
\end{figure}

In \cite{Benini:2009gi}, it was realised that the $E_7$ theory can be realised as $4$ M5-branes wrapped over a sphere with $3$ punctures. The punctures are of the type $SU(4)$, $SU(4)$, $SU(2)$. This theory is depicted in the left picture of \fref{f:TA3}.  The Lagrangian description of this theory is unknown.  

We denote the $E_7$ theory by a `quiver diagram' analogue to those in previous sections.  This is given in the right picture of \fref{f:TA3}.  The green blob denotes the theory with unknown Lagrangian description.  The $E_7$ global symmetry is indicated in the square node.

\begin{figure}[htbp]
\begin{center}
\includegraphics[totalheight=1.5cm]{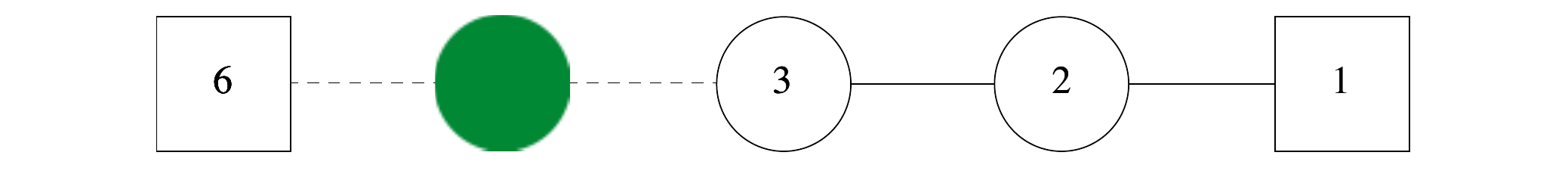}
\caption{{\bf The $6-\bullet-3-2-1$ quiver theory:}  The global symmetry $E_7$ can be decomposed into the subgroup $SU(3) \times SU(6)$.
The $SU(3)$ symmetry is gauged and is coupled to the $3-2-1$ tail. The $U(1)$ global symmetries are associated with the solid lines in the quiver diagram.
The global symmetry is thus $SU(6) \times U(1) \times U(1)$.}
\label{f:E7af}
\end{center}
\end{figure}

\begin{figure}[htbp]
\begin{center}
\includegraphics[totalheight=1.8cm]{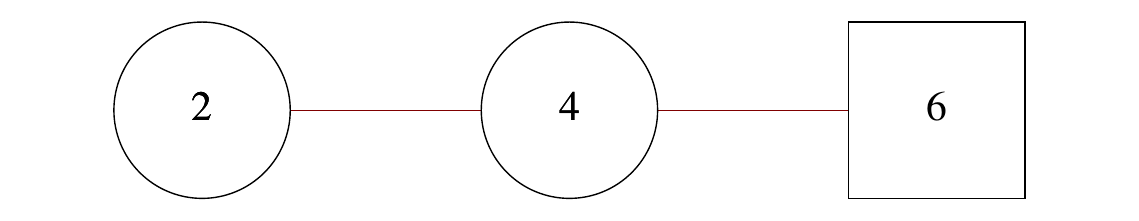}
\caption{{\bf The $2-4-6$ quiver theory.}  
This theory is dual to the $6-\bullet-3-2-1$ quiver theory.}
\label{f:246}
\end{center}
\end{figure}

The $E_7$ theory can be used to construct a quiver gauge theory called the $6-\bullet-3-2-1$ theory, depicted in \fref{f:E7af}.  The duality between this theory and the $2-4-6$ quiver theory (depicted in \fref{f:246}) is proposed by \cite{Benini:2009gi}.  Our purpose of this section is to construct and match the Hilbert series of both sides of the duality.  


Let us summarise a construction of the $6-\bullet-3-2-1$ quiver theory.  
The global symmetry $E_7$ can be decomposed into the subgroup $SU(3) \times SU(6)$.
The $SU(3)$ symmetry is gauged and is coupled to the $3-2-1$ tail, depicted in \fref{f:E7af}.
The $U(1)$ global symmetries are associated with the hypermultiplets and hence the solid lines in the quiver diagram.
The global symmetry is thus $SU(6) \times U(1) \times U(1)$.  

A trick to obtain the $3-2-1$ tail is to consider the $SU(2)$ theory with 4 flavours, whose flavour symmetry of is $SO(8)$.  The group $SO(8)$ contains $SU(4) \times U(1) \supset SU(3) \times U(1) \times U(1)$ as subgroups.  Gauging the $SU(3)$ group in $SO(8)$ and gluing it to the $SU(3)$ group in $E_7$, we obtain the $6-\bullet-3-2-1$ quiver theory.


On the other side of the duality,  we have the $2-4-6$ quiver theory, depicted in \fref{f:246}.  
The $U(1)$ global symmetries are associated with the hypermultiplets and hence the solid lines in the quiver diagram.
Therefore, the flavour symmetry is $U(6) \times U(1) \cong SU(6) \times U(1) \times U(1)$, in agreement with that of the $6-\bullet-3-2-1$ quiver theory.   
From the quiver diagram, it is clear that the $2-4-6$ quiver theory can also be obtained by gauging the $SU(2)$ subgroup of the $U(8)$ flavour group of the $SU(4)$ gauge theory with 8 flavours.

\subsubsection{The Hilbert series of the $2-4-6$ quiver theory}
In this subsection, the refined and unrefined Hilbert series are computed.  The former contains information about the global symmetries and how the gauge invariants transform under such symmetries, whereas the latter contains information about the dimension of the moduli space and the number of operators in the spectrum.  In order to compute an exact form of the refined Hilbert series, general formulas involving branching rules need to be determined.  However, such formulas can sometimes be very cumbersome and difficult to compute; in which case, what one can do is to compute the first few orders of the refined Hilbert series.  Nevertheless, it may be possible that the unrefined Hilbert series can be computed exactly.  We give an example below.

The $2-4-6$ quiver theory can be obtained by gauging the $SU(2)$ subgroup of the $U(8)$ flavour group of the $SU(4)$ gauge theory with 8 flavours. The Hilbert series written in terms of $SU(8)$ representations is given by  \eref{refu6}.  We first discuss a branching rule for $SU(8)$ to $U(1) \times SU(2) \times SU(6)$. 

\paragraph{A branching rule for $SU(8)$ to $U(1) \times SU(2) \times SU(6)$.} A map from the $SU(8)$ fugacities $x_1, \ldots, x_7$ to the $U(1)$ fugacity $q$, the $SU(2)$ fugacity $z$ and the $SU(6)$ fugacities $y_1, \ldots, y_5$ can be
\bea
\begin{array}{llll}
x_1= q y_1,  &\quad x_2 = q^{2} y_2, &  \quad x_3 = q^{3} y_3, &\quad x_4 = q^{4} y_4, \nn \\
 x_5 = q^{5} y_5, & \quad x_6 = q^{6}, & \quad x_7 = q^{3} z~. &
\end{array}
\eea 
For example, we have
\bea
\left[1,0,0,0,0,0,0 \right] &=& [0; 1,0,0,0,0] q + [1; 0,0,0,0,0] q^{-3} \nn \\
\left[1,0,0,0,0,0,1 \right] &=& [0;0,0,0,0,0]+[2;0,0,0,0,0]+[1;0,0,0,0,1]q^{-4} \nn \\
&& + [1;1,0,0,0,0]q^4+[0;1,0,0,0,1]~.
\eea
Using this decomposition, the Hilbert series of the $SU(4)$ theory with 8 flavours can be written as
{\small
\bea
&& \gH_{N_c = 4, N_f =8} = 1+ (2+[2;0,0,0,0,0]+[1;0,0,0,0,1] \frac{1}{q^4}+ [1;1,0,0,0,0] q^4 \nn \\
&& \quad + [0;1,0,0,0,1]) t^2 + \Big(4+2 [2;0,0,0,0,0]+[4;0,0,0,0,0]+\frac{3 [1;0,0,0,0,1]}{q^4} \nn \\
&& \quad +\frac{[3;0,0,0,0,1]}{q^4}+\frac{[2;0,0,0,0,2]}{q^8}+\frac{[0;0,0,0,1,0]}{q^8}+\frac{q^4 [0;0,0,0,1,0]}{b^2} \nn \\
&& \quad+b^2 q^4 [0;0,0,0,1,0]+\frac{[1;0,0,1,0,0]}{b^2}+b^2 [1;0,0,1,0,0]+\frac{[0;0,1,0,0,0]}{b^2 q^4} \nn \\
&& \quad+\frac{b^2 [0;0,1,0,0,0]}{q^4}+q^8 [0;0,1,0,0,0]+q^4 [1;0,1,0,0,1]+[0;0,1,0,1,0] \nn \\
&& \quad+3 q^4 [1;1,0,0,0,0]+q^4 [3;1,0,0,0,0]+3 [0;1,0,0,0,1]+2 [2;1,0,0,0,1] \nn \\
&& \quad+\frac{[1;1,0,0,0,2]}{q^4}+\frac{[1;1,0,0,1,0]}{q^4}+q^8 [2;2,0,0,0,0]+q^4 [1;2,0,0,0,1] \nn \\
&& \quad+[0;2,0,0,0,2] \Big) t^4+ \ldots~.
\eea}

\paragraph{The refined Hilbert series of the $2-4-6$ theory.}
This can be computed by gauging the $SU(2)$ symmetry. The gauging is done by integrating over the $SU(2)$ Haar measure and Supersymmetry imposes additional adjoint valued F terms, which are written below as the glue factor,
\beq\label{E7}
g_{2-4-6} (t;q;b; y_1, \ldots, y_5) =  \int \ud \mu_{SU(2)}  ~g_{\mathrm{glue}} ~ \gH_{N_c=4, N_f=8}~,
\eeq
where the gluing factor is given by
\bea
g_{\mathrm{glue}} (t; z) = \frac{1}{\PE\left[ [2]_{SU(2)} t^2 \right]} = 1 - [2]  t^2 +  [2] t^4 - t^6 ~.
\eea
The integral in \eref{E7} projects out the $SU(2)$ singlets.  This gives
\bea \label{su4su2ref}
&& g_{2-4-6} (t; q, b; y_1, \ldots, y_5) = 1+ (2+[1,0,0,0,1]) t^2 + \Big(3+\frac{1}{q^4}[0,0,0,1,0] \nn \\
&& \quad +\frac{q^2}{b^2}[0,0,0,1,0]+b^2 q^2[0,0,0,1,0]+\frac{1}{b^2q^2}[0,1,0,0,0]+\frac{b^2}{q^2}[0,1,0,0,0] \nn \\
&& \quad +q^4[0,1,0,0,0]+[0,1,0,1,0]+3[1,0,0,0,1]+[2,0,0,0,2]\Big) t^4 + \ldots ~.
\eea

\paragraph{The unrefined Hilbert series.} The unrefined Hilbert series can be computed exactly.  Setting $q=b=y_1=\ldots = y_5=1$ in \eref{E7}, it can be easily seen that the integrand is simply a rational function of $t$ and $z$. Evaluating the integral, one obtains the closed form
\bea
g_{2-4-6} (t) &=& \frac{P(t)}{(1- t^2)^{28}(1 + t^2)^{14}} \nn \\
&=&  1+37 t^2+792 t^4+12180 t^6+145838 t^8+1422490 t^{10} + \ldots~. \label{directsu4su2}
\eea
where 
\bea
P (t) &=&  1+23 t^2+351 t^4+3773 t^6+29904 t^8+180648 t^{10}+855350 t^{12}+ \nn \\
&& 3243202 t^{14}+10014534 t^{16}+25512281 t^{18}+54163863 t^{20}+ \nn \\
 && 96566265 t^{22}+ 145392195 t^{24}+185575556 t^{26}+201252816 t^{28}  \nn \\
 &&  + \ldots~\text{(palindrome)}~\ldots+ t^{56}~.
\eea
The plethystic logarithm of this Hilbert series is
\bea
\PL [g_{2-4-6} (t)] = 37 t^2+89 t^4-252 t^6-2800 t^8+14720 t^{10}+124524 t^{12} + \ldots~. \nn \\
\eea

\subsubsection{The Hilbert series of the $6-\bullet-3-2-1$ quiver theory}
As described in \sref{sec:duale7}, the $6-\bullet-3-2-1$ quiver theory can be obtained by `gluing' the $SU(3)$ subgroup of the $E_7$ theory with the $SU(3)$ subgroup of the $SO(8)$ flavor symmetry for $SU(2)$ with $4$ flavors.   The Hilbert series of the latter, written in terms of $U(4)$ representations, is given in Equation \eref{gH24}.  In order to gauge the $SU(3)$ subgroup, one needs to find a branching rule for $SU(4)$ to $U(1) \times SU(3)$.

\paragraph{A branching rule for $SU(4)$ to $U(1) \times SU(3)$.} A map from the $SU(4)$ fugacities $x_1, \ldots, x_3$ to the $U(1)$ fugacity $q$ and the $SU(3)$ fugacities $z_1, z_2$ can be
\bea
x_1 = \frac{z_1}{q}, \quad x_2 = \frac{z_2}{q^2}, \quad x_3 = \frac{1}{q^3}~.
\eea 
With this map, one can rewrite \eref{gH24} in terms of $SU(3)$ representations as
\bea
\gH_{3-2-1} &=& \frac{1}{1-t^2} \sum_{n_1=0}^\infty \sum_{n_2 = 0}^\infty \sum_{n_3 = 0}^\infty [n_1, n_2+n_3, n_1]_{SU(4)} t^{2n_1+2n_2+2n_3} b^{2n_2-2n_3}  \nn \\
&=&  \frac{1}{{(1-t^2)^2} } \sum_{n_1=0}^\infty \sum_{n_2 = 0}^\infty \sum_{n_3 = 0}^\infty q^{2n_1-2n_2} \frac{b^{-2 (n_1+n_2)}(1-b^{4 (1+n_1+n_2)})}{(1-b^4)} \times \nn \\
&&  \Big[  [n_1+n_3, n_2+n_3] + \sum_{n_4=0}^{n_3-1}  ( q^{-4n_3+4n_4} [n_1+n_3, n_2+n_4]     \nn \\
&& + q^{4n_3-4n_4} [n_1+n_4, n_2+n_3] ) \Big] t^{2(n_1+n_2+n_3)}~.
\eea

Since we need to gauge $SU(3) \subset E_7$, we also need to obtain the branching rule of $E_7$ representations to the subgroup $SU(3) \times SU(6)$.

\paragraph{Branching rule for $E_7$  to $SU(3)\times SU(6)$.} 
The branching rules can be obtained by matching the characters on both sides.  A map of the $E_7$ fugacities $u_1, \ldots, u_7$ to the $SU(3)$ fugacities $z_1, z_2$ and the $SU(6)$ fugacities $y_1, \ldots, y_5$ can be 
\bea
\label{E7toSU3SU6}
u_1&=& z_1 y_2, \quad u_2 = y_1 y_2, \quad u_3 = z_2 y_2^2, \quad u_4 = y_2^3, \quad 
 u_5 = \frac{y_2^2 y_3}{y_4}, \quad u_6 = \frac{y_2^2}{y_4}, \quad u_7 = \frac{y_2 y_5}{y_4}~. \nn \\
\eea
For example, the decompositions of $Adj^1$ and $Adj^2$ of $E_7$ are given below.  We use the notation $[a_1, a_2; b_1, \ldots, b_5]$ to denote the representations of $SU(3)\times SU(6)$.
{\small
\begin{eqnarray}
Adj^1 &=& [1,1; 0,0,0,0,0] + [1,0; 0,1,0,0,0] + [0,1; 0,0,0,1,0]  + [0,0;1,0,0,0,1] \nn \\
Adj^2 &=& [2, 2; 0, 0, 0, 0, 0] + [2, 0;0, 2, 0, 0, 0] + [0, 2; 0, 0, 0, 2, 0] + [0, 0; 2, 0, 0, 0, 2] \nn \\
&+&[2, 1; 0, 1, 0, 0, 0] + [1, 1; 0, 1, 0, 1, 0] + [0, 1; 1, 0, 0, 1, 1] + [2, 0;0, 0, 0, 1, 0]  \nn \\
&+&[1, 2; 0, 0, 0, 1, 0] + [1, 0; 1, 1, 0, 0, 1] + [1, 1; 0, 0, 0, 0, 0] + [0, 2; 0, 1, 0, 0, 0] \nn \\
&+&[1, 1; 1, 0, 0, 0, 1] + [1, 0;0, 1, 0, 0, 0] + [1, 0; 0, 0, 1, 0, 1] + [0, 0; 1, 0, 0, 0, 1] \nn \\
&+&[0, 0; 0, 1, 0, 1, 0] + [0, 1; 0, 0, 0, 1, 0] + [0, 1;1, 0, 1, 0, 0] + [0, 0; 0, 0, 0, 0, 0] ~. \nn \\
\end{eqnarray}}
The Hilbert series of the coherent component of the one $E_7$ instanton moduli space on $\BR^4$ after using the fugacity map Equation \eref{E7toSU3SU6} is
\bea
\HS_{E_7} (t; z_1,z_2; y_1, \ldots, y_5) = \sum_{k=0}^\infty Adj^k (z_1, z_2; y_1, \ldots, y_5) t^{2k}~. \label{hse7un}
\eea

\paragraph{Gluing process.}  We obtain the Hilbert series of the $6-\bullet-3-2-1$ quiver theory by using a similar `gluing technique' to Equation \eref{integ}:
\beq\label{E7integ}
g_{6-\bullet-3-2-1} (t; q, b; y_1, \ldots, y_5) =  \int \ud \mu_{SU(3)} ~\HS_{E_7}~g_{\mathrm{glue}}~ \gH_{3-2-1}~,
\eeq
where the gluing factor is given by the adjoint valued F terms,
\bea
g_{\mathrm{glue}} (t;z_1,z_2)= \frac{1}{\PE \left[ [1,1]_{SU(3)} t^2 \right]}~.
\eea
Therefore, we obtain
\bea
&& g_{6-\bullet-3-2-1} (t; q, b; y_1, \ldots, y_5) = 1+ (2+[1,0,0,0,1]) t^2 + \Big(3+\frac{1}{q^8}[0,0,0,1,0] \nn \\
&& \quad +\frac{q^4}{b^2}[0,0,0,1,0]+b^2 q^4[0,0,0,1,0]+\frac{1}{b^2q^4}[0,1,0,0,0]+\frac{b^2}{q^4}[0,1,0,0,0] \nn \\
&& \quad +q^8[0,1,0,0,0]+[0,1,0,1,0]+3[1,0,0,0,1]+[2,0,0,0,2]\Big) t^4 + \ldots ~,
\eea
in accordance with \eref{su4su2ref}, up to a rescaling of $q$ (which means simply that we use different units in counting charges):
\bea
g_{6-\bullet-3-2-1} (t; q, b; y_1, \ldots, y_5) = g_{2-4-6} (t; q^2, b; y_1, \ldots, y_5)~.
\eea
Unrefining $b=q=y_1 = \ldots = y_5 =1$, we obtain the unrefined Hilbert series up to the order $t^8$ as
\bea
g_{6-\bullet-3-2-1} (t) =  1+37 t^2+792 t^4+12180 t^6+145838 t^8 + \ldots~. \label{e7su3}
\eea
This is in agreement with \eref{directsu4su2}.  






\subsection{$E_8$}
The resummed Hilbert series for the coherent branch of one $E_8$ instanton is 
\bea
\HS_{E_8} (t; 1, \ldots, 1)=  \frac{P_{E_8} (t)}{(1-t^2)^{58}}~,
\eea
where the numerator is a palindromic polynomial of degree $58$:
\bea
P_{E_8} (t) &=& 1+190 t^2+14269 t^4+576213 t^6+14284732 t^8+234453749 t^{10}+ \nn \\
&& 2675683550 t^{12}+ 21972715186 t^{14}+133126452657 t^{16}+606326972328 t^{18} +\nn \\
&& 2105555153625 t^{20}+ 5634990969615 t^{22}+11714759112330 t^{24}+ \nn \\
&& 19025183027595 t^{26}+24223919026560 t^{28}+ \ldots~\text{(palindrome)}~\ldots+ t^{58}~. \nn \\
\eea
This is consistent with the fact that the Higgs branch is $2h_{E_8}-2=58$ complex dimensional, where $h_{E_8}=30$ is the dual Coxeter number of $E_8$.

\begin{figure}[htbp]
\begin{center}
\includegraphics[totalheight=5cm]{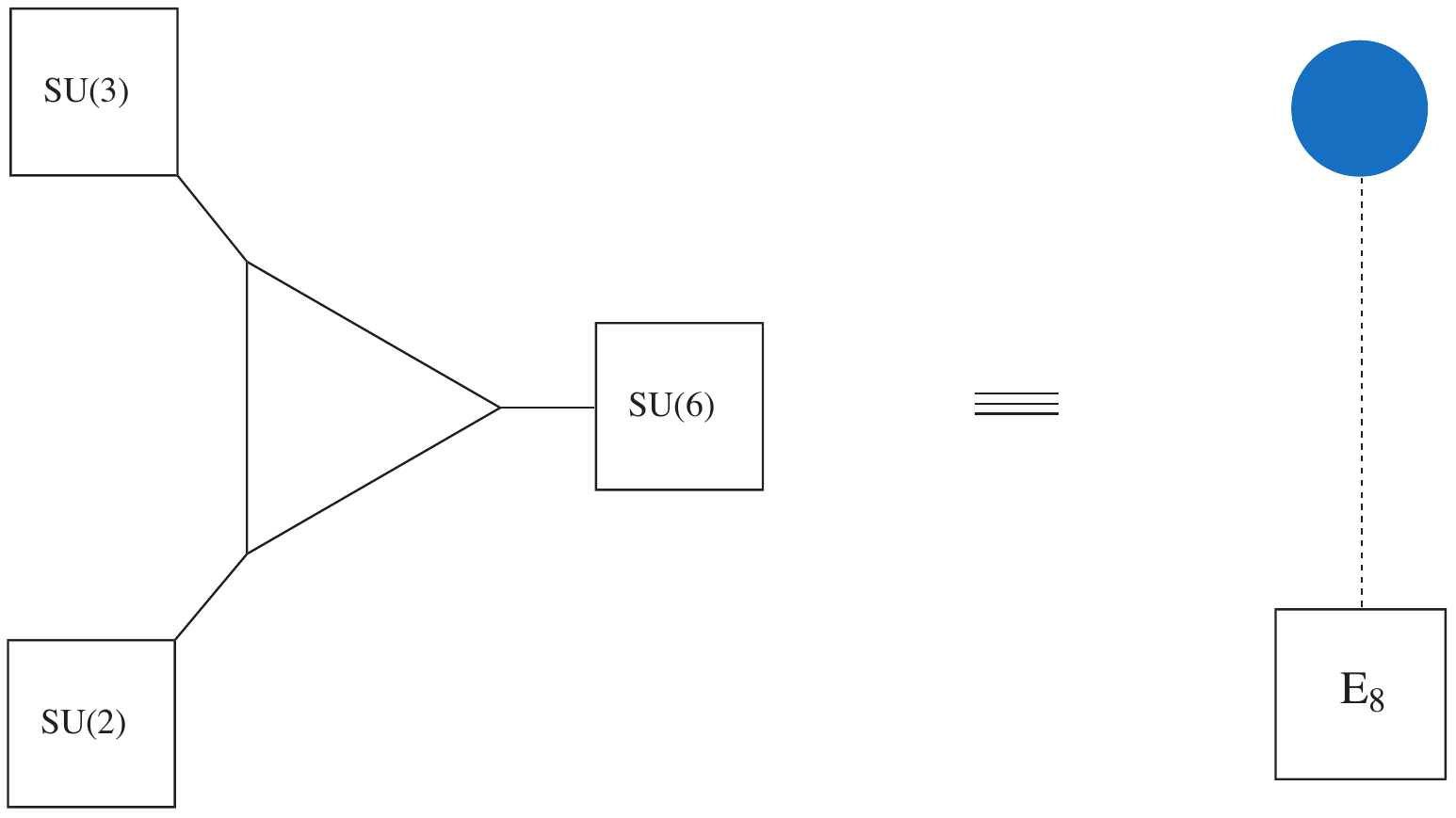}
\caption{{\bf Left:}  The $E_8$ theory arises from $6$ M5-branes wrapping a sphere with $3$ punctures. The $3$ punctures are of the type $SU(6)$, $SU(3)$, $SU(2)$. {\bf Right:}  The quiver diagram representing the $E_8$ theory.  The blue blob denotes a theory with an unknown Lagrangian description.  The $E_8$ global symmetry is indicated in the square node.}
\label{f:E8}
\end{center}
\end{figure}

The $E_8$ theory arises from $6$ M5-branes wrapping a sphere with $3$ punctures. The $3$ punctures are of the type $SU(6)$, $SU(3)$, $SU(2)$.  The quiver diagram is depicted in the left picture of \fref{f:E8}.  The Lagrangian description of this theory is unknown.  

We denote the $E_8$ theory by a `quiver diagram' analogue to those in previous sections.  This is given in the right picture of \fref{f:E8}.  The blue blob denotes a theory with an unknown Lagrangian description.  The $E_8$ global symmetry is indicated in the square node.

\begin{figure}[htbp]
\begin{center}
\includegraphics[totalheight=1.2cm]{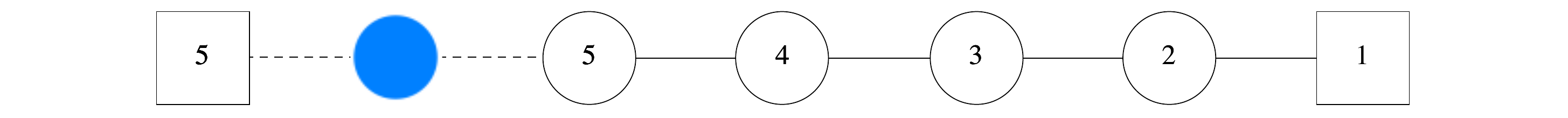}
\caption{The $5-\bullet-5-4-3-2-1$ quiver theory.  
The $U(1)$ global symmetries are associated with the solid lines in the quiver diagram.
The flavour symmetry is expected to be $SU(5) \times U(1)^4$.}
\label{f:E8be}
\end{center}
\end{figure}

\begin{figure}[htbp]
\begin{center}
\includegraphics[totalheight=3cm]{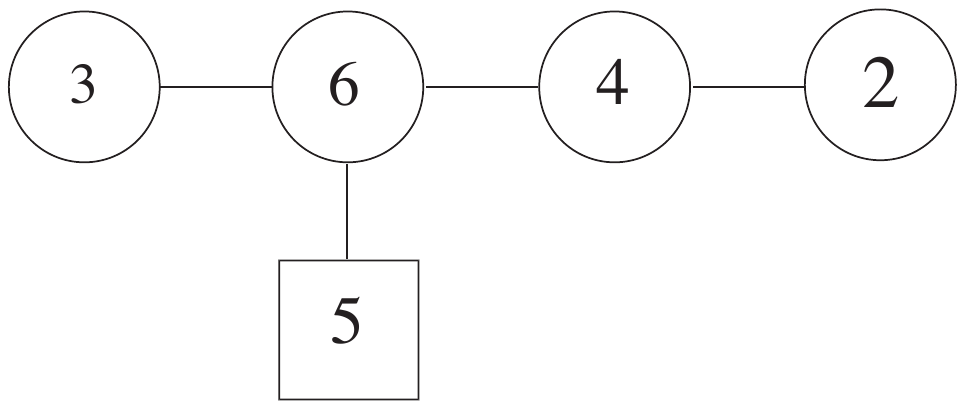}
\caption{The $3-6_{[5]}-4-2$ quiver theory. This theory is  dual to the $5-\bullet-5-4-3-2-1$ theory.}
\label{f:E8af}
\end{center}
\end{figure}

The $E_8$ theory can be used to construct a quiver gauge theory called the $5-\bullet-5-4-3-2-1$ theory, depicted in \fref{f:E8be}.  
The duality between this theory and the $3-6_{[5]}-4-2$ quiver theory (depicted in \fref{f:E8af}) is proposed by \cite{Benini:2009gi}.

The $5-\bullet-5-4-3-2-1$ theory can be constructed as follows.  The global symmetry $E_8$ can be decomposed into $SU(5) \times SU(5)$.  One of the $SU(5)$ is gauged and is coupled to the $5-4-3-2-1$ tail.  The $U(1)$ global symmetries are associated with the solid lines in the quiver diagram.
Hence, the flavour symmetry is expected to be $SU(5) \times U(1)^4$. 


On the other side of the duality, we have the $3-6_{[5]}-4-2$ quiver theory depicted in \fref{f:E8af}.  As in all previous quivers, the $U(1)$ global symmetries are associated with the solid lines in the quiver diagram, and the flavour symmetry is expected to be $
SU(5) \times U(1)^4$, in agreement with that of the $5-\bullet-5-4-3-2-1$ quiver theory. 



The computations of Hilbert series of these theories are rather involved and technical.  We leave such computations for future work.

\subsection{One $F_4$ instanton on $\BC^2$}
There is no simple analog of the ADHM construction. Instead the conjecture of this paper is that the Hilbert series for the one instanton moduli space on $\BC^2$ is a sum over symmetric adjoint representations. Explicitly, denote the adjoint representation of $F_4$ by $[1,0,0,0],$ and the symmetric adjoints by $[k,0,0,0]$, then the dimension of each representation is
\bea
&&  \dim~[k,0,0,0] = \\
& & \frac{(k+1) (k+2) (k+3)^2 (k+4)^3 (k+5)^2 (k+6) (k+7) (2 k+5) (2 k+7) (2
   k+9) (2 k+11)}{4191264000}, \nonumber
\eea
and the Hilbert series for the moduli space takes the form
\beq
g_{F_4}(t; x_1, x_2, x_3, x_4, x) = \frac{1}{(1-t x)(1-t/x)}\sum_{k=0}^\infty [k, 0, 0, 0] t^{2k}~,
\eeq
Where as usual, the first term is the Hilbert series for $\BC^2$, physically interpreted as the position of the instanton and the remaining function is the Hilbert series for the coherent component of the moduli space.
By setting the $F_4$ fugacities to 1 one can get an explicit palindromic rational function for the coherent component of the moduli space,
\beq
\HS_{F_4}(t) = \frac{1+36 t^2+341 t^4+1208 t^6+1820 t^8+1208 t^{10}+341 t^{12}+36t^{14}+t^{16}}{(1-t^2)^{16}}
\eeq
giving a non-trivial check that the dimension of this moduli space is $2(h-1)=16$, where $h=9$ is the dual Coxeter number of $F_4$.

\subsection{One $G_2$ instanton on $\BC^2$}
This case also has no known simple ADHM construction. Denote the character of the adjoint representation by $[0,1]$ and the character for the $k$-th symmetric adjoint by $[0,k]$, with dimension
\beq
\dim~[0,k] = \frac{(k+1) (k+2) (2k+3) (3k+4) (3k+5)}{120}~.
\eeq
The Hilbert series takes the form
\beq
g_{G_2}(t; x_1, x_2, x) = \frac{1}{(1-t x)(1-t/x)}\sum_{k=0}^\infty [0, k] t^{2k},
\eeq
and setting the fugacities to 1 gives
\beq
g_{G_2}(t; 1, 1, 1) = \frac{1}{(1-t)^2}\frac{1+8t^2+8t^4+t^6}{(1-t^2)^6}~,
\eeq
giving a non-trivial check that the dimension of this moduli space is $2(h_{G_2}-1)=6$, where $h_{G_2}=4$ is the dual Coxeter number of $G_2$.
Since the rank of this gauge group is 2, it is possible to compute the sum explicitly and write the Hilbert series as a rational function with characters of $G_2$. Omitting the trivial $\BC^2$ part we get
\beq
\HS_{G_2}(t; x_1, x_2) = P_{G_2}(t; x_1, x_2) \PE \left [ [0,1] t^2\right ]~,
\eeq
where $P_{G_2}$ is a palindromic polynomial of degree 11 in $t^2$ and has the form
\bea
P_{G_2}(t; x_1, x_2) &=& \nn
1 - ([2, 0] + 1) t^4 + ([1, 1] + [2, 0] + [0, 1]) t^6 - ([3, 0] + [1, 1] + [0, 1] + [1, 0]) t^8 \\ \nn
&+& ([3, 0] + [1, 0]) t^{10} + ([3, 0] + [1, 0]) t^{12} - ([3, 0] + [1, 1] + [0, 1] + [1, 0]) t^{14} \\
&+& ([1, 1] + [2, 0] + [0, 1]) t^8 - ([2, 0] + 1) t^{18} + t^{22}~.
\eea

\acknowledgments
We are indebted to Francesco Benini and Alberto Zaffaroni for useful discussions and to Sam Kitchen for his generosity in helping us in programming.  A.~H.~ would like to thank Ecole Polytechnique for their kind hospitality during the completion of this paper, and to Michael Douglas and Nikita Nekrasov for useful discussions.  N.~M.~ acknowledges Giuseppe Torri for a close collaboration and thanks John Davey, Ben Hoare and David Weir of their kind help.  He is grateful to the following institutes and collaborators for their kind hospitality during the completion of this work: Max-Planck-Institut f\"ur Physik (Werner-Heisenberg-Institut); Rudolf Peierls Centre for Theoretical Physics, University of Oxford; DAMTP, University of Cambridge; Universiteit van Amsterdam; Frederik Beaujean, Francis Dolan, Yang-Hui He, Sven Krippendorf and Alexander Shannon.  He also thanks his family for the warm encouragement and support. This research is supported by the DPST project, the Royal Thai Government.

\end{document}